\documentclass[lettersize,journal,10pt]{IEEEtran}
\IEEEoverridecommandlockouts
\usepackage{amsmath,amssymb,amsfonts}
\usepackage{algorithmic}
\usepackage[ruled,vlined,linesnumbered]{algorithm2e}
\usepackage{amsmath}
\usepackage{graphicx}
\usepackage{textcomp}
\usepackage{xcolor}
\usepackage{balance}
\usepackage{multirow}
\usepackage{threeparttable}
\usepackage{comment}
\usepackage[misc]{ifsym}

\ifCLASSOPTIONcompsoc
  \usepackage[nocompress]{cite}
\else
  \usepackage{cite}
\fi

\newcommand{\biographyspace}{\vspace{-10mm}}

\usepackage{dblfloatfix} % This is used to position the table more freely

\usepackage[caption=false,font=footnotesize]{subfig}

\def\BibTeX{{\rm B\kern-.05em{\sc i\kern-.025em b}\kern-.08em
    T\kern-.1667em\lower.7ex\hbox{E}\kern-.125emX}}

\begin{document}

\makeatletter
\def\ps@IEEEtitlepagestyle{%
  \def\@oddhead{%
    \parbox[b]{\textwidth}{\centering
      \fontsize{7}{8}\selectfont\itshape
      This work has been submitted to the IEEE for possible publication.
      Copyright may be transferred without notice, after which this
      version may no longer be accessible.}%
    \hfil}%
  \def\@evenhead{\@oddhead}%
  \def\@oddfoot{}%
  \def\@evenfoot{}%
}
\makeatother

%\title{STF-Less Packet Transmission in DECT-2020 NR: Overhead-Performance Trade-off Analysis}

% \title{DECT NR+ Topology Construction Algorithm for Ultra-Dense Deployments}

\title{Topology Construction Algorithms for DECT NR+ in Normal and Ultra-Dense Deployments}

\begin{comment}

\author{
\IEEEauthorblockN{
Andrey Samuylov$^{1}$, Dmitri Moltchanov$^{1}$, Juho Pirskanen$^2$, Jussi Numminen$^2$, Mikko Valkama$^{1}$}

\IEEEauthorblockA{
$^{1}$Unit of Electrical Engineering, Tampere University, Finland\\$^2$ Wirepas Oy, Tampere, Finland\\
\Letter \, Contact author's e-mail: roman.glazkov@tuni.fi
}\vspace{-0mm}}

\end{comment}

\author{Andrey Samuylov, Dmitri Moltchanov, Roman Glazkov, Juho Pirskanen, Jussi Numminen, Mikko Valkama

\thanks{A. Samuylov, D. Moltchanov, R. Glazkov, and M. Valkama are with Tampere Wireless Research Center, Tampere University, Finland. E-mail: firstname.lastname@tuni.fi}
\thanks{J. Pirskanen and J. Numminen are with Wirepas Oy, Tampere, Finland. E-mail: firstname.lastname@wirepas.com}}

% Faculty of Information Technology and Communication Sciences,

%%%%%%%%%%%%%%%%%%%%%%%%%%%%%%%%%%%%%%%%%%%%%%%%%%%%
\maketitle
%%%%%%%%%%%%%%%%%%%%%%%%%%%%%%%%%%%%%%%%%%%%%%%%%%%%
\begin{abstract}
The 5G-grade DECT NR+ technology designed to support massive machine-type communications (mMTC) relies on the decentralized mesh connectivity concept. The currently utilized topology construction algorithms are very sensitive to the node density, which causes interference at the node association phase and may lead to instable topologies. In this study, we propose several algorithms that need to be utilized at relay and leaf nodes to allow stabilization of the topology: (i) load-dependent relay allocation, (ii) hysteresis-based association control, (iii) blacklist-based association control, and (iv) smart beacon placement. Due to the complex interference structure in DECT NR+ mesh systems, we utilize system-level simulations to assess the performance in both normal and ultra-dense conditions in terms of the efficiency of topology construction and the performance characteristics of the resulting topology. Our results show that dense DECT NR+ networks do suffer from topology instability, with reassociation rates exceeding 10 per second and recovery times lasting over 10 minutes. By implementing the proposed mechanisms, the network dynamically scales, reducing the number of required relays by more than half, from 34\% to 15\%. They also stabilize the topology by reducing the reassociation rates to just 2--3 per second allowing large-scale networks of several hundred nodes to operate efficiently. %Finally, we show that load-dependent relay allocation is a critical mechanism that needs to be implemented in the DECT NR+ system to ensure that the network can adapt to its size dynamically.
%\textcolor{magenta}{Our numerical results show that  the proposed approach optimizes the network structure by reducing the number of active relay nodes from 50\% to 20\% and increasing stability 20-fold by lowering reassociation events from 40 events/s to just 2 events/s. Furthermore, the proposed solution improves the overall efficiency by providing a 2.25x reduction in the average hop count while maintaining a high packet delivery ratio between 0.95 and 0.98.} %it reduces the number of relays in the system while still keeping all the nodes connected. At the data transmission stage, the resulting topology provides performance comparable to non-dense systems, allowing the utilization of DECT NR+ in new application areas characterized by very good propagation conditions, such as agriculture and drone smarms.
\end{abstract}
%%%%%%%%%%%%%%%%%%%%
%%%%%%%%%%%%%%%%%%%%

\section{Introduction}\label{sect:intro}

% DECT: standardized, utilized for mMTC/URLLC

DECT New Radio (NR+) is a recently standardized technology that operates in the unlicensed band \cite{x1}. It has recently been accepted by ITU-R as a full-grade 5G-compliant standard for massive machine-type communications (mMTC), supporting 1 million user equipment (UE) devices per square kilometer \cite{kovalchukov2022dect} while satisfying the mMTC service requirements in \cite{itur_m_2410} by delivering 99\% of packets within the 10 s delay budget. Recently, DECT NR+ has been shown to satisfy much stricter latency and packet delivery requirements that approach those of ultra-reliable low-latency service (URLLC) \cite{samuylov2025empowering,glazkov2025improving}.

% Toploogy constriction is critical

Similarly to other multi-hop systems such as Integrated Access and Backhaul (IAB) or unmanned aerial vehicle (UAV) networks \cite{fiore2025meeting,gaydamaka2023dynamic}, topology construction is a critical feature of DECT NR+ systems that affects not only the principal connectivity performance but also the performance metrics delivered by the system at the data transmission phase. As detailed in Section \ref{sect:related}, performance of the DECT NR+ topology construction procedure has received little attention thus far. Studies that addressed this challenge concentrated on low-to-moderate densities of UEs in the network and/or on indoor deployments, where propagation is heavily obstructed by walls and other obstacles in the channel. Consequently, the interference domain of a single UE was small, causing no issues during the topology construction phase. This is especially important in the case of blackouts or mass installations, when all nodes attempt to establish a topology almost instantaneously.

% Dense topologies

When ultra-dense deployments caused by extreme UE density or by excellent propagation conditions that are close to free-space path loss (FSPL) typical for, e.g., drone swarms or open-area use cases are considered, not only does the interference domain of a single UE become very crowded, causing excessive interference already at the beaconing phase, but also the potential choices for associations are abundant. This causes significant challenges for the distributed UE association and relay selection algorithms, which are essential for DECT NR+ topology construction. Specifically, inefficient topology configurations caused by associations with UEs located far away and subsequent reassociations mixed with varying interference conditions may lead to constantly changing topologies.

% What we do

The aim of this study is to develop algorithms to improve the topology construction in DECT NR+ systems suitable for both normal and ultra-dense deployment conditions. To this end, we propose four enhancement algorithms that together improve the UE association and relay selection procedure, namely: (i) load-dependent relay allocation, (ii) hysteresis-based association control, (iii) blacklist-based association control, and (iv) smart beacon placement. The first technique relaxes the restrictive assumption on the maximum number of nodes that can be associated with a single relay node. The next two mechanisms slow down the association such that a better path in the network can be found, while the fourth mechanism reduces interference during the beaconing phase. The proposed algorithms do not change the Medium Access Control (MAC) procedures specified in the DECT NR+ standards; they only add logic on top of them. To capture the complex reassociation and interference dynamics of the transient period, we utilize system-level simulations to investigate the system's behavior. We evaluate not only topology-related metrics, including association and reassociation rates, and the number of hops to the sink, but also the eventual performance of the resulting topology in terms of packet delivery ratio (PDR), mean latency, and the 95th percentile of latency.

The main contributions of this study are as follows:
\begin{itemize}
    \item{A set of algorithmic improvements to conventional DECT NR+ UE association and relay selection mechanisms specifically designed for DECT NR+ systems in ultra-dense deployments or environments with exceptionally good propagation (e.g., airborne systems).}
    \item{An implementation framework in a system-level simulation showing that the proposed solutions require no modifications to the current DECT NR+ standards.}
    \item{Numerical results showing that: (i) load-dependent relay allocation enables DECT NR+ networks to adapt seamlessly to varying network sizes, (ii) without optimization, dense networks suffer chaotic reconfigurations with reassociation rates exceeding 10 events/s, (iii) the proposed mechanisms stabilize the network, maintaining reassociation rates under 3 events/s in systems with up to 700 nodes, (iv) active relays in the system are cut by more than half, dropping from 34\% to just 15\% of the overall number of nodes, (v) the solutions prevent severe drops in PDR and delay, avoiding system time-to-steady-state periods exceeding 10 min.}
    %, (vi) eliminating artificial interference allows stable operation of hundreds of nodes in ultra-dense or airborne scenarios
    %  from 40 events/s to just 2 events per second for a 600-node network
    % for a 2.25x reduction in the average number of hops while maintaining a high packet delivery ratio (PDR) of 0.95–0.98
\end{itemize}

% How the paper is organized

The remainder of this paper is organized as follows. In Section \ref{sect:related}, we briefly review the principles of DECT NR+ technology and outline the related work. In Section \ref{sect:appr}, we describe the conventional UE association and relay selection algorithms and propose our algorithms. The system model and evaluation methodology are outlined in Section \ref{sect:eval}. The numerical results are presented in Section \ref{sect:num}. The conclusions are drawn in the final section.

\section{Background and Related Work}\label{sect:related}

In this section, we first provide a brief description of the DECT NR+ operational principles. Then, we sketch the related work done in the field. 

\subsection{DECT NR+ Systems}

\begin{figure}[t!]
    \centering
    \includegraphics[width=0.9\columnwidth]{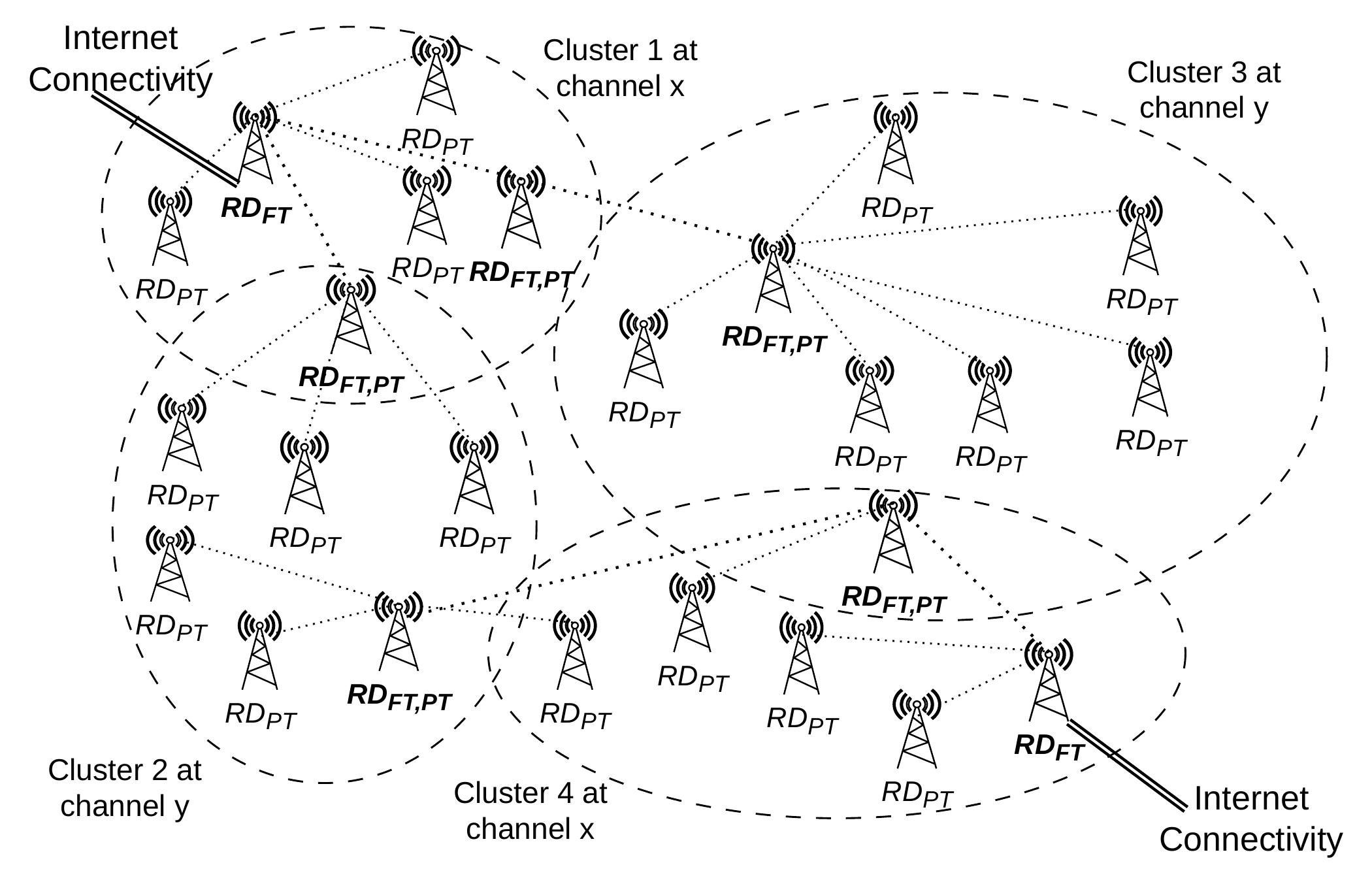}
    \caption{Illustration of DECT NR+ mesh network architecture.}
    \label{fig:meshIllus}
    \vspace{-4mm}
\end{figure}

To support mesh-based IoT deployments, DECT NR+ structures its network using a cluster tree topology, see Fig. \ref{fig:meshIllus}, allowing radio devices (RDs) to act as fixed terminations (FT) and portable terminations (PT) \cite{x1,samuylov2023performance}. Upon initialization, RDs evaluate the available link qualities and route costs, which measure the path efficiency to external backends, to select an optimal FT node for association. Active FT RDs serve as routers that coordinate local radio resources and anchor incoming PT connections to the network. They relay data from connected PTs, along with their own local traffic, until they reach a network sink that interfaces with external systems, such as the Internet.

At the physical layer, DECT NR+ leverages Cyclic Prefix Orthogonal Frequency Division Multiplexing (CP-OFDM) \cite{x2}. It inherits several advanced 5G NR capabilities, including variable subcarrier spacing, hybrid automatic repeat request (HARQ), and adaptive modulation and coding schemes (MCSs). However, unlike the 3GPP 5G NR, the DECT NR+ PHY layer operates strictly below 6 GHz, relies solely on time-division duplexing (TDD), and supports multiple-input multiple-output (MIMO) operation with up to eight streams.

The MAC layer in DECT NR+ relies on a listen-before-talk (LBT) protocol with a binary exponential backoff (BEB) scheme for uplink transmissions across both modes. For random access channel (RACH) transmissions, the LBT sensing window spans only two symbols, ensuring low-latency medium access \cite{dect_samuylov_sim_estim}. For downlink distribution, the architecture supports mechanisms such as cached downlink routing and selective flooding \cite{TS103636-5}.

\subsection{Related Work}

% System-level

Recent studies on DECT NR+ systems confirm that the technology is a 5G-grade multi-hop solution designed for mMTC and near-URLLC services. Specifically, the research performed so far has addressed the capacity of these systems under the mMTC traffic assumptions specified in ITU-R M.2412 \cite{itur_m_2412} and the mMTC service requirements specified in ITU-R M.2410 \cite{kovalchukov2022dect,anttonen2021enabling}. These studies have shown that the system capacity over a single channel scales to well above one million UEs per square kilometer, with each UE generating a single 32 KB packet every 2 hours. A recent study in \cite{samuylov2025empowering} demonstrated that DECT NR+ can also support near-URLLC applications characterized by latencies on the order of 10--50 ms and a packet delivery ratio approaching $0.99$. Support for special applications, such as industrial deployments \cite{wassmann2024performance} and high-fidelity audio \cite{durre2024dect,poets2025professional}, has been evaluated.

% Link layer

The studies performed so far have also addressed the physical layer performance of various transmission modes for both mMTC \cite{penner2021link} and URLLC applications \cite{penner2021urllc}. Recently, a study assessing link- and system-level performance jointly has been reported \cite{Bandara2026Impact}. Specifically, it has been shown that the use of receiver diversity may provide a simple and cost-efficient solution for enhanced mMTC performance.

% Deployments

A set of studies considered the performance of DECT NR+ systems under various deployment conditions. In contrast to the indoor-hall and mixed indoor-hall/outdoor deployments considered in previous studies, the authors of \cite{poets2025professional} evaluated DECT NR+ as a candidate technology for professional audio applications across the 450, 1350, and 3700 MHz frequency bands, while the study in \cite{wassmann2024performance} addressed an industrial deployment. The experimental results in the latter study showed that the DECT NR+ system could achieve an average one-way latency of 1.7 ms, which is significantly lower than the latencies reported for standard Wi-Fi or 5G NR solutions. Thus, this technology is a promising candidate for the stringent technical demands of the Programme Making and Special Events (PMSE) sector. The authors of \cite{poets2025professional} demonstrated that DECT NR+ outperforms classic DECT in Rayleigh fading channels while also showing that lower frequencies, specifically 450 MHz, offer superior reliability with packet loss below 5\%. Additionally, the study utilized OFDM-based channel sounding to characterize the industrial indoor wireless channel, finding an average root-mean-square (RMS) delay spread of 65 ns for a 13.824 MHz bandwidth.

% Topology construction and maintenance

Although DECT NR+ is based on multi-hop operation, almost no studies conducted so far have addressed topology construction procedures. The exact algorithmic implementation details of the beaconing and scanning procedures that enable topology construction are specified in ETSI TS 103 636-4 \cite{x4}. The only exception is  \cite{Glazkov2026Improving}, where the authors investigated the topology construction algorithm of the DECT NR+ standard and identified that the current static approach can lead to unstable network configurations during initialization or after blackouts. To address these instabilities, the authors propose two enhancements, dynamic load-dependent relay selections and hysteresis-based association control, designed to ensure sufficient relay availability and prevent nodes from reassociating frequently owing to occasional interference. The numerical results demonstrate that these improvements can halve the time required for the network to reach full connectivity, reduce the time-to-steady-state for data transmission by 40\%, and improve the mean packet delay by 10\%.

% Problem

One critical limitation of multi-hop systems is that they perform poorly in ultra-dense network conditions, where there are numerous UEs in the interference domain of a single node. During the data transmission phase, potential collisions can be avoided by utilizing the LBT operation. However, the greatest adverse impact occurs during association and relay selection, where the system operates according to a slotted ALOHA. The aim of this study is to enable an efficient topology construction algorithm for such deployment conditions.

\section{The Proposed Approach}\label{sect:appr}

In this section, we first describe the currently utilized UE association and relay selection procedures in DECT NR+ systems. Subsequently, we describe the proposed mechanisms.

%\subsection{Association and Relay Section Procedures}

\begin{figure}[t!]
    \centering
    \includegraphics[width=0.55\columnwidth]{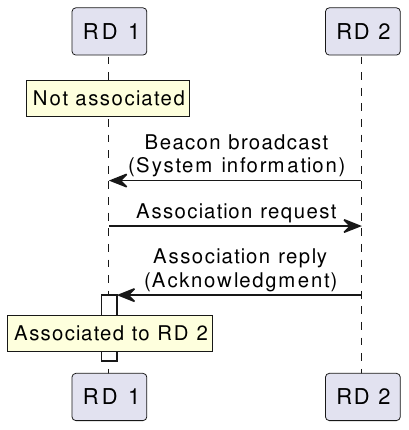}
    \caption{Illustration of the association mechanism in DECT NR+.}
    \label{fig:association_sequence}
    \vspace{-4mm}
\end{figure}

\subsection{Association and Relay Selection in DECT NR+}

% At the glance approach

Topology construction in multi-hop systems plays a crucial role in the overall system performance. For an individual RD, it consists of three steps: (i) UE association, (ii) role (PT or FT) selection, and (iii) beaconing. First, during network initialization or recovery after a blackout, each RD performs a scanning process to identify the available neighboring nodes. Once a beacon with a Received Signal Strength Indicator (RSSI) higher than a given threshold is sensed, the RD proceeds with the association procedure. After that, an RD chooses whether to operate in the PT mode (as a leaf node) or in the FT mode. In FT mode, the device acts as a relay/router, coordinating local resources and allowing other PT RDs to connect. In the latter case, the FT-mode RD selects an appropriate time instant for its beacons and starts broadcasting them.

% Associations

\subsubsection{UE Association Procedure} 

A generic association signaling procedure in DECT NR+ is defined in \cite{x4} and is shown in Fig. \ref{fig:association_sequence}. An RD willing to establish or change an association first scans the environment for nearby RDs operating in FT mode by listening to their broadcast beacon messages. Beacon information is collected over the observation window, allowing the RD to form a set of viable candidate FTs.

A candidate is considered viable when its received beacon strength satisfies the minimum link-quality requirement and the candidate has available association capacity. The RD then compares the viable candidates using the route-selection criteria, including path length toward the sink and signal-to-noise ratio (SNR). Once a candidate is selected, the RD sends an association request on the RACH using the access parameters announced in the beacon. The RD waits for the association response and, if the request is accepted, confirms the new association and releases the previous association, if one exists. If the request is rejected or no response is received within the fixed response window, the association attempt is considered unsuccessful. The RD may then continue the association process by evaluating another viable candidate or by waiting for a later association opportunity.
\subsubsection{Relay Selection}

For scalability, RDs can independently select their operational mode, either FT or PT. An RD operating in PT mode may switch to FT mode after establishing a connection to an RD already operating in the FT mode, thereby enabling it to serve as a relay for additional RDs. The design of the relay selection mechanism must consider multiple factors, including limited RD power resources, overall network connectivity and traffic performance requirements.

Mode-switching decisions are governed by a set of scenario- and quality-of-service (QoS)-dependent rules. Before transitioning to the FT mode, an RD must confirm that it is connected to an uplink RD in the FT mode and then monitor its radio environment for a defined observation period to ensure that the number of neighboring FT-mode RDs does not exceed a specified threshold. This threshold is a configurable parameter selected to ensure continuous network connectivity. Only neighboring FT-mode RDs with link quality above a minimum threshold determined by the target MCS are considered. The observation period is chosen to reliably capture at least one beacon from the neighboring FT-mode RDs.

\begin{figure}[!t]
    \centering
    \includegraphics[width=0.3\columnwidth]{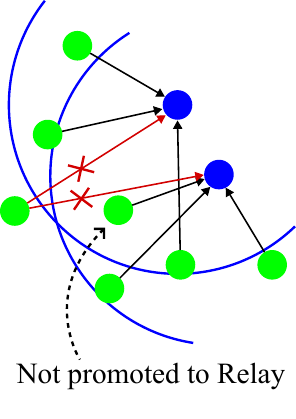}~~~~~~~
    \includegraphics[width=0.34\columnwidth]{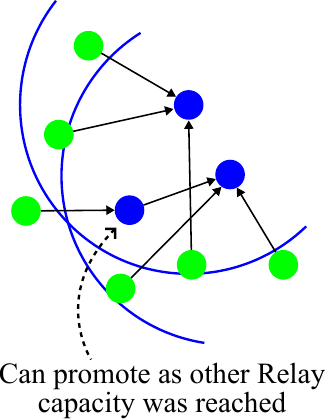}
    \caption{Illustration of load-dependent relay selection.}
    \label{fig:load-dep-illus}
    \vspace{-4mm}
\end{figure}

\subsubsection{Limitations of Existing Approaches}

There are several areas where DECT NR+ UE association and relay-selection mechanisms need to be extended to allow for ultra-dense deployment conditions. The conventional algorithm operations in ultra-dense environments face the following challenges:
\begin{itemize}
    \item{\textit{Limited relay availability.} The conventional DECT NR+ relay selection does not account for the number of nearby nodes when selecting an RD as a relay. Even in normal deployments, where local density of nodes can be generally low, local fluctuations in the number of nodes may require dynamic selection of RDs operating in FT mode.}
    \item{\textit{Short-term association instability.} This challenge is caused by the high number of UEs available for connectivity in close propagation proximity to a given UE. The chances that a UE located far away will be chosen for connectivity are relatively high; thus, there will be a number of times when the beacons from closer FTs will be heard, causing reassociations. This problem is further complicated by dynamic interference that is caused by both beaconing and packet transmission for those UEs that have already found a route to the sink.}
    \item{\textit{Long-term association instability.} This challenge is closely associated with the previous one, as the availability of nodes with slightly better propagation conditions and the same or a lower number of hops to the sink causes switches between next hops increasing the number of association messages sent. However, it is critical to allow the UE to choose better routes to the sink, as it generally improves the topology. Thus, two algorithms are needed: one suppressing the short-term UE churn, while the second freezing the node for a longer time.}
    \item{\textit{Random beacon transmission times.} By default, the beacon transmission time is chosen randomly at each RD that has been selected to operate as FT. In dense conditions, when the number of neighbors can reach several hundred, this causes beacon overlap during the topology construction phase increasing the likelihood of UE reassociation due to beacons being suppressed by interference.}
\end{itemize}

Finally, there is a critical point related to the lack of an explicit boundary between the topology construction and data transmission phases in DECT NR+. In fact, once the path towards the sink is available, nodes may start transmitting traffic. This coexistence of associations and data transmission may negatively affect both phases.

\subsection{Load-Dependent Relay Selection}

\begin{algorithm}[t!]
    \KwIn{Count threshold $T_{\text{th}}$, occupied-capacity threshold $\gamma$, minimum scan duration $T_{\text{beacon}}$}
    \KwOut{Updated operational mode $M \in \{\text{PT},\text{FT}\}$}

    \While{RD is active}{
        Perform environment scan for $t \ge T_{\text{beacon}}$\;
        Store the set of viable association candidates $\mathcal{C}$\;
        $N_{\text{cand}} \leftarrow |\mathcal{C}|$\; %\tcp*{Candidates}

        \If{$N_{\text{cand}} < T_{\text{th}}$}{
            $M \leftarrow \text{FT}$\; %\tcp*{Switch to FT mode}
        }
        \Else{
            Determine occupied association capacity $S_{\text{used}}$ and
            total association capacity $S_{\text{total}}$ from the
            broadcast beacons\;

            \If{$S_{\text{total}} = 0$}{
                $\rho \leftarrow 1$\;
            }
            \Else{
                $\rho \leftarrow S_{\text{used}} / S_{\text{total}}$\;
            }

            \If{$\rho > \gamma$}{
                $M \leftarrow \text{FT}$\; %\tcp*{Switch to FT mode}
            }
            \Else{
                Maintain current mode\;
            }
        }
    }
    \caption{Load-Dependent Relay Selection}
    \label{alg:load_dep_relay}
\end{algorithm}

The first point to address is \textit{limited relay availability}, ensuring that the already associated RDs can propagate the network coverage further, improving the overall network connectivity as long as the geometrical locations of UEs allow it. To enable this, RDs can switch their operational mode to FT when needed and start broadcasting beacon messages.

The baseline approach is to scan the environment for other RDs available for association and switch the operational mode from PT to FT if their number is lower than a given threshold. The shortcoming of this approach is that a constant threshold may not be suitable for all deployments. To avoid this, we introduce a load-dependent relay selection algorithm that supplements the candidate-count threshold with an estimate of local association capacity shown in Fig. \ref{fig:load-dep-illus}. An associated RD periodically scans the environment for at least one beacon interval and stores all viable association target candidates, $N_{\text{cand}} = |\mathcal{C}|$. Let $S_{\text{used}}$ and $S_{\text{total}}$ be the total occupied association capacity reported by neighboring FTs and the total association capacity reported by those FTs, respectively. The occupied-load fraction is then defined as 
\begin{align}
\rho=\begin{cases}
S_{\text{used}}/S_{\text{total}} & \text{if }  S_{\text{total}}>0,  \\
1 &  \text{if }  S_{\text{total}}=0.\\
\end{cases}
\end{align}

That is, the operational mode, $M \in \{\text{PT}, \text{FT}\}$, is determined dynamically using the occupied-load fraction $\rho$ and the preset free-slot fraction threshold $\gamma$ and the present candidated threshold $T_{\text{th}}$ as
\begin{equation}
        M = \begin{cases} 
        \text{FT}, & \text{if } N_{\text{cand}} < T_{\text{th}}, \\ 
        \text{FT}, & \text{if } \rho > \gamma, \\ 
        \text{PT}, & \text{otherwise}.
        \end{cases}
    \end{equation}

In contrast to the conventional DECT NR+ approach, this scheme results in the dynamic selection of relays depending on the current load condition in the vicinity of a node. The associated algorithm is shown in Algorithm \ref{alg:load_dep_relay}. %The algorithm is provided in Algorithm \ref{alg:load_dep_relay}. %Building on the load-dependent relay selection and hysteresis-based association control introduced in \cite{Glazkov2026Improving}, the main novelty of this study is the extension of the association and relay selection procedure to ultra-dense and near-FSPL deployments through blacklist-based association control and smart beacon placement, where complex interference patterns arising from mesh-based connectivity may otherwise result in unpredictable network behavior. %Note that this approach is parameterized by a single parameter -- the local load conditions. \textcolor{red}{What is that?}

\subsection{Hysteresis and Bias-Based Association Control}

\begin{algorithm}[t!]
    \KwIn{Current FT $FT_{\text{curr}}$, current RSSI $R_{\text{curr}}$, current hop count $H_{\text{curr}}$, set of candidate FTs $\mathcal{F}$, minimum RSSI threshold $R_{\text{min}}$, hysteresis margin $h$}
    \KwOut{Selected upstream FT $FT_{\text{selected}}$}
    
    $FT_{\text{selected}} \leftarrow FT_{\text{curr}}$\;
    $R_{\text{best}} \leftarrow R_{\text{curr}}$\;
    $H_{\text{best}} \leftarrow H_{\text{curr}}$\;
    
    \ForEach{candidate $FT_j \in \mathcal{F}$ with RSSI $R_j$ and hop count $H_j$}{
        \If{$R_j \ge R_{\text{min}}$}{
            \If{$H_j < H_{\text{best}}$}{
                $FT_{\text{selected}} \leftarrow FT_j$\;
                $R_{\text{best}} \leftarrow R_j$\;
                $H_{\text{best}} \leftarrow H_j$\;
            }
            \ElseIf{$H_j = H_{\text{best}}$ \textbf{and} $R_j > R_{\text{best}} + h$}{
                $FT_{\text{selected}} \leftarrow FT_j$\;
                $R_{\text{best}} \leftarrow R_j$\;
                $H_{\text{best}} \leftarrow H_j$\;
            }
        }
    }
    \caption{Bias and Hysteresis Association Control}
    \label{alg:association_control}
\end{algorithm}

The second mechanism addresses the \textit{short-term association instability} of UE association decisions in dense DECT NR+ topologies. In the conventional procedure, an RD may observe several FT candidates with comparable hop counts and similar received beacon powers. Under ultra-dense conditions, small RSSI variations caused by fading, beacon interference, or packet transmissions may trigger frequent changes in the selected upstream FT. This behavior increases signaling overhead and may disturb the topology construction process itself.

To suppress these effects, we utilize association control based on received-power bias and hysteresis, a technique widely used in cellular systems \cite{karmakar2022mobility,lee2008adaptive}. The bias component prevents the UE from considering beacon sources whose RSSI is only marginally above the sensitivity region, thereby reducing the probability of selecting weak upstream links. The hysteresis component further requires a candidate FT to provide a sufficiently higher RSSI before replacing the currently selected FT when the hop counts toward the sink are comparable. Candidates that provide a shorter path to the sink may still be preferred so that the mechanism does not freeze the topology completely. Instead, it filters out slight link quality fluctuations while preserving the ability of RDs to switch to better association targets.

Let $FT_{\text{curr}}$ denote the currently associated FT with received signal strength $R_{\text{curr}}$ and hop count $H_{\text{curr}}$. Let $\mathcal{F}$ be the set of discovered candidate FTs, where each candidate $FT_j \in \mathcal{F}$ has an RSSI $R_j$ and a hop count $H_j$. To prevent connections to marginally acceptable links, a candidate is filtered out unless its RSSI satisfies a minimum power threshold $R_{\text{min}}$ incorporating a received-power bias above the sensitivity limit,
\begin{align}
R_j \ge R_{\text{min}}.
\end{align}

After the viable candidates have been observed, the candidate FT $FT_j$ that replaces the current FT $FT_{\text{curr}}$ must satisfy either a shorter-path criterion or an equal-hop-count criterion combined with a hysteresis margin $h$:
\begin{align}
\hspace{-3mm}\text{Switch to } FT_j \iff
        \begin{cases} 
        H_j < H_{\text{curr}}, \\ 
        H_j = H_{\text{curr}} \text{ and } R_j > R_{\text{curr}} + h,  
        \end{cases}\hspace{-4mm}
\end{align}
where the former branch accounts for the smaller hop count, while the latter -- for the same hop count and hysteresis.

The associated algorithm is shown in Algorithm \ref{alg:association_control}.

\subsection{Smart Beacon Randomization}

\begin{figure}[t!]
    \centering
    \includegraphics[width=0.8\columnwidth]{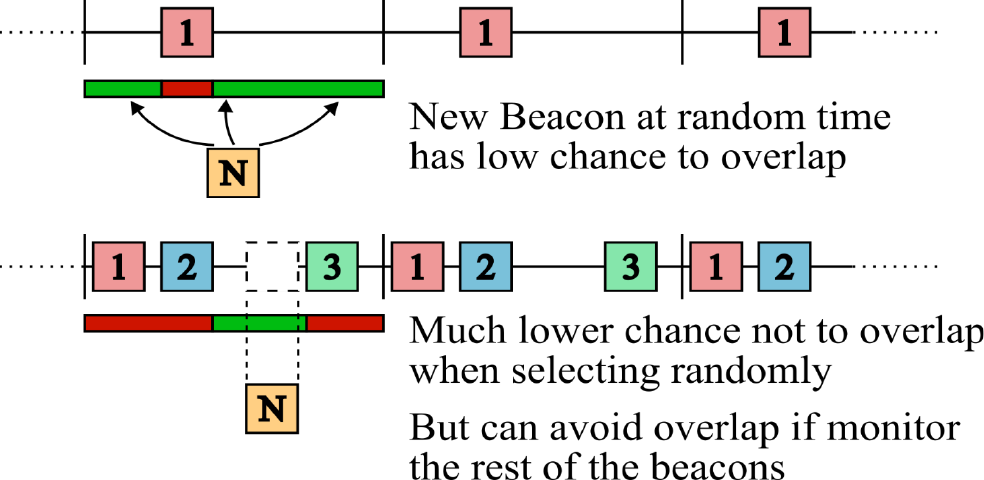}
    \caption{The conventional and proposed beaconing approaches.}
    \label{fig:beaconing}
    \vspace{-4mm}
\end{figure}

The third mechanism targets unintended synchronization and beacon overlaps during topology construction addressing the problem of \textit{random beacon transmission times}. When many RDs start operating nearly simultaneously, as may happen after a blackout or during mass deployment, association searches and FT beacon transmissions may become temporally aligned. In a dense interference domain, this produces bursts of random-access attempts and repeated beacon overlaps. Consequently, some UEs may miss otherwise suitable FT candidates, select suboptimal parents, or trigger additional reassociations after the missed beacons become observable again.

Smart randomization spreads both association search activity and FT beacon transmissions in time, as shown in Fig. \ref{fig:beaconing}. Randomized association start times reduce simultaneous RACH access, whereas randomized beacon start times reduce alignment among neighboring FTs. In addition, beacon placement can be made locally conflict-aware: an FT-mode RD can select a beacon position that avoids transmission windows already observed from nearby FTs. The resulting effect is reduced beacon interference, lower association contention, and fewer reassociation waves caused by transient beacon loss.

% Randomized start times

To mitigate simultaneous random-access attempts after initialization, an associating device delays its association search by a random jitter offset $\Delta t_{\text{assoc}}$, drawn from a uniform distribution over a contention window $W_{\text{assoc}}$,
\begin{align}
\Delta t_{\text{assoc}} \sim \mathcal{U}(0, W_{\text{assoc}}),
\end{align}
and thus, the association start time is defined as $T_{\text{assoc}} = t_0 + \Delta t_{\text{assoc}}$, where $t_0$ is the system boot or wake-up time.

% Conflict-aware placement

To introduce conflict-aware beacon placement, let $\mathcal{B}_{\text{obs}} = \{t_1,t_2,\dots,t_n\}$ denote the most recent transmission timestamps observed from neighboring FTs within listening range. Each observed beacon also provides its transmission period through the beacon information received and maintained by the observing RD. The local FT uses its own configured beacon interval $T_{\text{local}}$ to define the candidate transmission cycle. This local interval is not inferred from neighboring beacon observations. Instead, the observed neighbor periods are used to predict the recurring transmission windows of the neighboring beacons. Let $T_i$ denote the advertised transmission period associated with neighbor beacon timestamp $t_i$.

An FT-mode RD selects its beacon transmission time $T_{\text{beacon}}$ from a set of candidate positions $\mathcal{T}_{\text{cand}}$ within its local beacon cycle. The selected position maximizes the minimum temporal distance, or guard band $\Delta g$, from the predicted neighboring beacon windows:
\begin{align}
    &T_{\text{beacon}} =
        \arg\max_{t \in \mathcal{T}_{\text{cand}}}
        \left[
            \min_i
            \left\{
                \min
                \left(
                    \phi_i(t), T_i - \phi_i(t)
                \right)
            \right\}
        \right],\nonumber\\
&\phi_i(t) = (t-t_i) \pmod{T_i}.
\end{align}

The associated algorithm is provided in Algorithm \ref{alg:beacon_randomization}.

\begin{algorithm}[t!]
    \KwIn{Association window $W_{\text{assoc}}$, beacon jitter window $W_{\text{beacon}}$, local beacon interval $T_{\text{local}}$, and observed neighbor beacon information $\mathcal{B}_{\text{obs}}$}
    \KwOut{Scheduled association start time $T_{\text{assoc}}$ and scheduled beacon transmission time $T_{\text{beacon}}$}

    \textbf{Phase 1: Randomized association start} \\

    Generate random association delay $\Delta t_{\text{assoc}} \leftarrow \mathcal{U}(0,W_{\text{assoc}})$\;

    Set $T_{\text{assoc}} \leftarrow t_{\text{current}}+\Delta t_{\text{assoc}}$\;

    \If{RD operates in FT mode}{
        \textbf{Phase 2: Conflict-aware beacon placement} \\

        Observe neighboring beacon transmissions for at least one local beacon cycle and collect their latest timestamps and advertised periods\;

        Define the candidate set $\mathcal{T}_{\text{cand}}$ within the local beacon cycle of duration $T_{\text{local}}$\;

        Initialize selected beacon position $T_{\text{beacon}} \leftarrow \emptyset$\; Initialize maximum minimum-distance $D_{\text{max}} \leftarrow -1$\;

        \ForEach{candidate position $t \in \mathcal{T}_{\text{cand}}$}{
            Set $d_{\min} \leftarrow +\infty$\;

            \ForEach{observed neighbor beacon $i$}{
                Calculate its predicted phase at position $t$ using its advertised period $T_i$\;

                Calculate the temporal distance $d_i$ from $t$ to the nearest edge of the neighbor's beacon transmission window\;

                $d_{\min} \leftarrow \min(d_{\min},d_i)$\;
            }

            \If{$d_{\min} > D_{\text{max}}$}{
                $D_{\text{max}} \leftarrow d_{\min}$\;
                $T_{\text{beacon}} \leftarrow t$\;
            }
        }

        \If{no suitable candidate position is available}{
            Generate a random beacon jitter $\Delta t_{\text{beacon}} \leftarrow \mathcal{U}(0,W_{\text{beacon}})$\;

            Set $T_{\text{beacon}} \leftarrow t_{\text{base}}+\Delta t_{\text{beacon}}$\;
        }
    }

    \caption{Beacon Randomization and Placement}
    \label{alg:beacon_randomization}
\end{algorithm}

\subsection{Association Target Blacklisting}

\begin{figure}[t!]
    \centering
    \includegraphics[width=0.5\columnwidth]{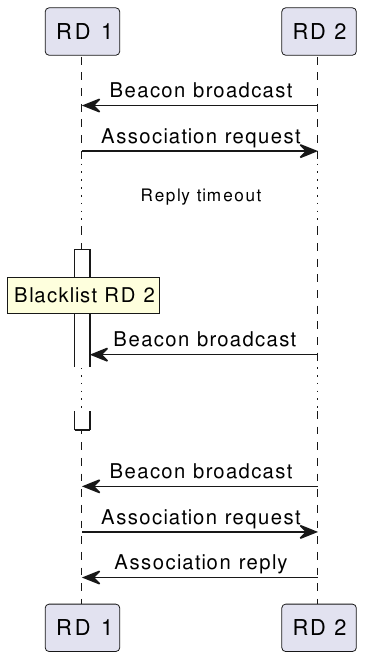}
    \caption{Illustration of blacklist operation.}
    \label{fig:blacklist-illus}
    \vspace{-4mm}
\end{figure}

The fourth mechanism addresses \textit{long-term association instability} by temporarily excluding association targets that recently resulted in unsuccessful or undesirable association behavior, as shown in Fig. \ref{fig:blacklist-illus}. Hysteresis alone cannot be made arbitrarily strong as UEs should remain able to discover better routes. Therefore, blacklisting complements hysteresis by slowing down repeated attempts to associate with problematic FT candidates while allowing these candidates to become available again after the transient condition has passed.% the transient condition has passed.

The blacklist is activated by several classes of events. First, stale target information is detected when a selected FT is no longer observable during association or deassociation signaling, or when an existing upstream FT disappears from the observed beacon set. Second, unanswered association or renewal attempts indicate that the target may be unreachable, congested, or temporarily hidden by interference. Third, an explicit rejection indicates that the target FT is reachable but does not currently accept the requesting RD. Finally, after a successful voluntary handoff, the previous FT may be temporarily avoided to prevent an immediate reassociation. These cases jointly reduce repeated signaling toward unsuitable targets and decrease long-term topology oscillations while permitting reselection when radio and load conditions change.

\begin{algorithm}[t!]
    \KwIn{Candidate set $\mathcal{F}$, blacklist $\mathcal{BL}$, current time $t_{\text{current}}$, blacklisting duration $T_{\text{dur}}$}
    % \KwIn{Candidate set $\mathcal{F}$, blacklist $\mathcal{BL}$, current time $t_{\text{current}}$, blacklisting duration $T_{\text{default\_dur}}$, block duration $T_{\text{HO\_dur}}$}
    \KwOut{Valid candidates $\mathcal{F}_{\text{valid}}$, updated blacklist $\mathcal{BL}$}
    
    \textbf{Step 1: Expire old blacklist entries} \\
    \ForEach{entry $(FT_i, T_{\text{exp}}(FT_i)) \in \mathcal{BL}$}{
        \If{$t_{\text{current}} \ge T_{\text{exp}}(FT_i)$}{
            Remove $FT_i$ from $\mathcal{BL}$\;
        }
    }

    \textbf{Step 2: Evaluate trigger events and update blacklist} \\
    \If{Selected $FT_j$ disappears before association transmission}{
        $\mathcal{BL} \leftarrow \mathcal{BL} \cup \{(FT_j,t_{\text{current}}+T_{\text{dur}})\}$\;
    }
    \If{Association request times out or $FT_j$ rejects association}{
        $\mathcal{BL} \leftarrow \mathcal{BL} \cup \{(FT_j,t_{\text{current}}+T_{\text{dur}})\}$\;
    }
    \If{Associated $FT_j$ is stale, $FT_j \notin \mathcal{B}_{\text{obs}}$, or activity timeout is exceeded}{
        $\mathcal{BL} \leftarrow \mathcal{BL} \cup \{(FT_j,t_{\text{current}}+T_{\text{dur}})\}$\;
    }
    \If{Renewal request is explicitly rejected}{
        $\mathcal{BL} \leftarrow \mathcal{BL} \cup \{(FT_j,t_{\text{current}}+T_{\text{dur}})\}$\;
    }
    \If{Association changes from $FT_{\text{prev}}$ to another FT}{
        $\mathcal{BL} \leftarrow \mathcal{BL} \cup \{(FT_{\text{prev}},t_{\text{current}}+T_{\text{dur}})\}$\;
    }

    \textbf{Step 3: Filter candidate set} \\
    $\mathcal{F}_{\text{valid}} \leftarrow \emptyset$\;
    \ForEach{Candidate $FT_j \in \mathcal{F}$}{
        \If{$FT_j \notin \mathcal{BL}$}{
            $\mathcal{F}_{\text{valid}} \leftarrow \mathcal{F}_{\text{valid}} \cup \{FT_j\}$\;
        }
    }
    
    \caption{Association Target Blacklisting}
    \label{alg:blacklisting}
\end{algorithm}

Let $\mathcal{BL}$ represent the active blacklist database. Each entry corresponds to a blacklisted FT identifier $FT_i$, associated with an expiration timestamp $T_{\text{exp}}(FT_i)$:
\begin{align}
\mathcal{BL} = \{ (FT_i, T_{\text{exp}}(FT_i)) \mid t_{\text{current}} < T_{\text{exp}}(FT_i) \}.
\end{align}

An FT candidate $FT_j$ is added to the blacklist $\mathcal{BL}$ when an association or maintenance event indicates that the target is temporarily unavailable or unreliable. This occurs in the following cases: (i) $FT_j$ disappears before an association or release message can be transmitted, (ii) an association request times out, (iii) the target explicitly rejects an association or renewal request, (iv) an association target becomes stale or is no longer present in $\mathcal{B}_{\text{obs}}$, or (v) the activity timeout indicates a lost association. After a successful handoff, the previous FT $FT_{\text{prev}}$ is also temporarily blacklisted to prevent an immediate return to the previous target. Each blacklist entry is assigned an expiration
timestamp:
\begin{align}
    T_{\text{exp}}(FT_j) = t_{\text{current}} + T_{\text{dur}}.
\end{align}

During target selection, the set of viable candidates $\mathcal{F}$ is filtered into a valid selection pool $\mathcal{F}_{\text{valid}}$ by excluding any target present in the active blacklist:
\begin{align}
\mathcal{F}_{\text{valid}} = \{ FT_j \in \mathcal{F} \mid FT_j \notin \mathcal{BL} \}.
\end{align}

The associated algorithm is shown in Algorithm \ref{alg:blacklisting}.

\section{System and Evaluation}\label{sect:eval}

In this section, we first describe the tool utilized for the DECT NR+ performance assessment and outline the modeling approach. Then, we introduce the considered deployments and radio part models. Finally, we define the metrics of interest and describe the data collection method.

\subsection{Simulation Tool and Modeling Approach}

The DECT NR+ system was implemented using WINTERsim, a discrete-event system-level simulation tool that played a pivotal role in the official standardization process of the technology \cite{kovalchukov2022dect}. In this environment, RDs are represented by a node abstraction class that encapsulates physical parameters such as spatial coordinates, antenna patterns, and power configurations while managing the entire functional stack, including the MAC layer, signaling engines, and traffic generators. To maintain computational efficiency in ultra-dense deployment scenario, the simulation utilizes a physical-layer abstraction that tracks bidirectional path-loss states, selectively processing only those signals and interference components that exceed the system's noise floor.

The physical layer is implemented according to DECT NR+ specifications, incorporating key functionalities such as HARQ and MCSs. To isolate the effects under study, several physical-layer features were intentionally enabled or disabled in the different scenarios considered in Section \ref{sect:num}, including the energy-saving sleep mechanism and power control. At the MAC layer, we focus exclusively on uplink transmissions from PT nodes to the sink and explicitly model the LBT procedure using BEB. The simulation parameters and configurations are summarized in Table \ref{tab:simScen}.

\subsection{Considered Deployment}

\begin{figure}[t!]
    \centering
    \includegraphics[width=0.5\columnwidth]{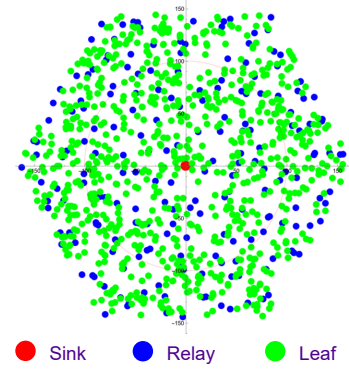}
    \caption{Typical node and relay locations in the considered deployment.}
    \label{fig:deploy}
    \vspace{-4mm}
\end{figure}

The ultra-dense scenario can be modeled by either increasing the density of nodes per unit area or by considering FSPL propagation model as in both cases the number of neighbors in the propagation proximity of RD increases. We adopt the latter approach as it allows to maintain the same node density and compare systems with normal and ultra-dense conditions under the same load conditions at each node.

We assume a network configuration with a single sink positioned at the origin of the coordinate system, as shown in Fig. \ref{fig:deploy}. The system operates at a carrier frequency of 1.9 GHz using one channel with a bandwidth of 1.728 MHz, and the network coverage area is limited to a radius of 170 m. Routers are not preassigned and they are dynamically elected according to the relay selection procedures outlined previously. To capture a range of operating conditions, the total number of nodes in the network was varied from 200 to 700. Each node was equipped with an independent traffic source that generated packet bursts consisting of two packets, with the time between successive bursts following an exponential distribution.

\subsection{Propagation and Antenna Models}

To compare the performance of the proposed approach, we considered two deployments with different propagation conditions: (i) FSPL and (ii) a 3GPP-based mixed indoor-hall/urban model. The former model is fully described by the Friis transmission equation and captures deployments with unobstructed propagation conditions, such as communications between unmanned aerial vehicles (UAVs). 

The second propagation model, representing standard conditions, is based on a set of 3GPP-defined propagation models for the 1.9 GHz band, as specified in 3GPP TR 38.901 \cite{3gpp_tr_38_901}. Depending on the node position, a propagation model is chosen as follows: (i) if one of the nodes is a sink, then the UMa path loss model is utilized; (ii) otherwise, if $d_{2D} < 25$ m, where $d_{2D}$ is the horizontal projection distance between the interacting nodes, then the InH path loss model is used; and (iii) in the remaining cases, the UMi model is employed. 

In addition, we account for line-of-sight (LoS) and non-line-of-sight (NLoS) conditions using the standardized LoS blockage probabilities in \cite{3gpp_tr_38_901}. For the UMa/UMi models, 
\begin{align}\label{eq:uma_los}
&p_{L}^{(U)} =
\begin{cases}
1, & d_{2D} \leq 18 \\
\frac{18}{d_{2D}} + e^{- \frac{d_{2D}}{36}} (1 - \frac{18}{d_{2D}}), & 18 < d_{2D} \\
\end{cases},
\end{align}
and for the InH model,
\begin{align}
&p_{L}^{(I)} =
\begin{cases}
1, & d_{2D} \leq 1.2 \\
e^{- \frac{d_{2D} - 1.2}{4.7}}, & 1.2 < d_{2D} < 6.5 \\
0.32 e^{- \frac{d_{2D} - 6.5}{32.6}}, & 6.5 \leq d_{2D} \\
\end{cases}.
\end{align}

Finally, we accounted for additional building penetration losses for the UMa and UMi models if at least one node is indoors. Another loss factor was added if both nodes are indoors and $d_{2D} > 25$ m. The outdoor-to-indoor building penetration loss for each indoor node is drawn independently as recommended in \cite{3gpp_tr_38_901}, i.e.,
\begin{align}\label{eq:penetration_loss}
P_{open} = 20 + 0.5\, U(0, 25),
\end{align}
where $U(0,x)$ is uniformly distributed in $[0,x]$.

\begin{table}[t!]
    \centering
    \caption{Considered system and evaluation parameters.}
    \label{tab:simScen}
    \begin{center}
        \bgroup
        \def\arraystretch{1.0}
        \begin{tabular}{p{0.5\columnwidth}|p{0.4\columnwidth}}
            \hline
            \multicolumn{1}{c|}{\textbf{Parameter}} & \multicolumn{1}{c}{\textbf{Value}} \\
            \hline\hline
            DECT NR+ FT sink & 1, height 25 m, outdoors \\ \hline
            Carrier frequency & 1900 MHz \\ \hline
            Number of RDs & 200--700 \\ \hline
            Number of channels & 1 \\ \hline
            Channel bandwidth & 1.728 MHz \\ \hline
            RD deployment & Uniformly random \\ \hline
            Building penetration loss & According to \eqref{eq:penetration_loss} \\ \hline
            Indoor/outdoor fraction & 80\% / 20\%\\ \hline
            Channel model, Sink-RD & Urban macro \\ \hline
            Channel model, RD-RD & Urban street canyon \\ \hline
            Number of RD/sink antenna elements & 1 \\ \hline
            Sink/RD noise figure & 7 dB \\ \hline
            Sink/RD antenna element gain & 0 dBi \\ \hline
            Thermal noise level & -174 dBm/Hz \\ \hline
            Traffic model & Bursts of 2 packets \\ \hline
            Application data message size & 32 bytes \\ \hline
            Mean interpacket time & 10 s \\ \hline
            Interpacket time distribution & Exponential \\ \hline
            Channel access & LBT + BEB \\ \hline
            Contention window size & min -- 24, max -- 768 \\ \hline
            LBT threshold & -75 dBm \\ \hline
            Minimum link quality & 15 dB \\ \hline
            Error correction & HARQ with 3 attempts \\
            \hline
            Sensitivity & -99.7 dBm (TS 103.636-2) \\ \hline
            RD Tx power & 10 dBm \\ \hline
            Format & MCSs 1/3/4 for 1/2/3 packets \\ \hline
            Default hysteresis value & 5 dB \\ \hline
            Beaconing period & 1 s \\ \hline
            Blacklist duration & 10 s \\ \hline
        \end{tabular}
        \egroup
    \end{center}
    \vspace{-4mm}
\end{table} 

%\subsection{Antenna Models}

% Propagation and antenna models

Owing to the low complexity of the RDs and the sink, we assume single-element antennas for all communicating devices with omnidirectional radiation patterns, operating in transmission mode (TM) 1, as specified in \cite{x2}. The receive and transmit gains and the noise floor are selected according to \cite{x2}, see Table \ref{tab:simScen}. 

%\subsection{Simulation Details and Procedure}

\subsection{Metrics and Data Collection}

% Metrics

To evaluate the performance of the topology construction algorithm, we consider two distinct categories of metrics. The first category focuses on the topology construction phase and comprises: (i) total number of associated nodes by time $t$, (ii) the association rate at time $t$, and (iii) the reassociation rate at time $t$. The former metric characterized the connectivity properties of the utilized algorithms and their convergence time. The instantaneous association rate indicates how quickly the network topology is formed and becomes operational for data transmission. In contrast, the reassociation rate provides insight into the temporal stability of the constructed topology. Note that for mesh systems, the reassociation rate should not be zero for the nodes to discover better paths to the sink. However, it should not be so high that signaling traffic overloads the network and creates additional interference. We also report on structural properties of the resulting topology using the mean number of hops to the sink and mean number of active relays in the system as two metrics of interest.

The second category of metrics characterizes the efficiency of the data transmission. These performance indicators include the mean packet delay, the 95th percentile of packet delay, and PDR, which together offer a view of latency and reliability during the data communication phase.

% Data collections

The evaluation of identified metrics requires the use of transient simulations. To obtain statistically meaningful results, we adopted an independent replication approach, as described in \cite{perros2009computer}. Each replication was generated using a different random seed, and for every configuration, ten independent runs were conducted. The identification of the steady-state onset was achieved by monitoring the reassociation rate using exponentially weighted moving average (EWMA) statistics. The system is deemed to have reached steady-state once variations in the EWMA of the reassociation rate fall below 0.01 events per second.

\begin{figure}[!t]
    \vspace{-0mm}
    \centering
    \subfloat[Without load-dependent relay allocation]{%
        \includegraphics[width=0.32\textwidth]{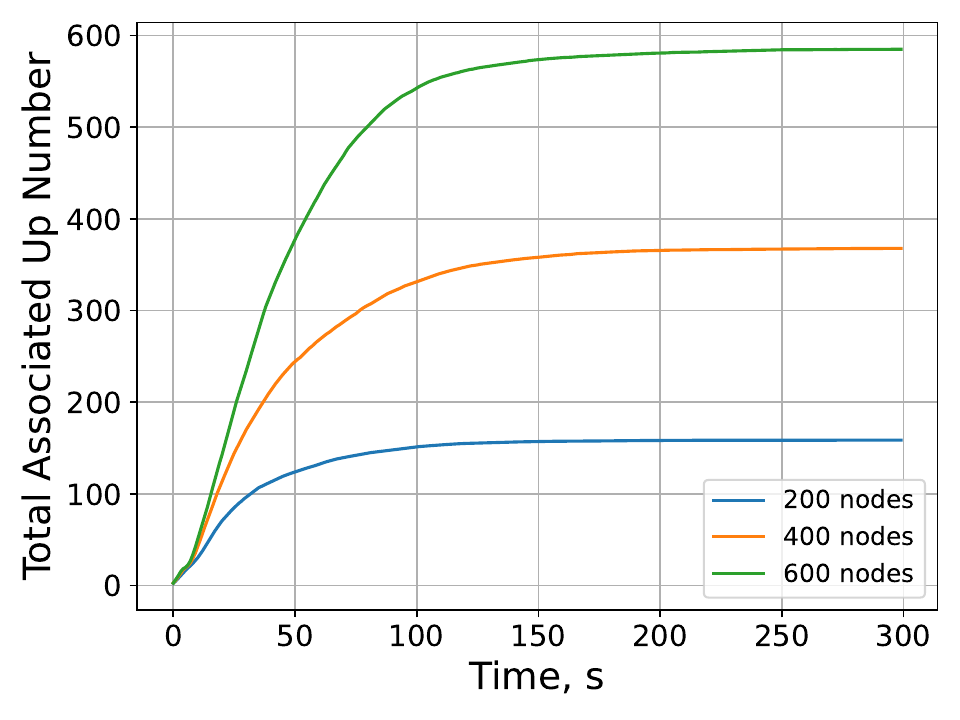}%
        \label{fig:totalRelay_inh_without}
    }\hfill
    \subfloat[With load-dependent relay allocation]{%
        \includegraphics[width=0.32\textwidth]{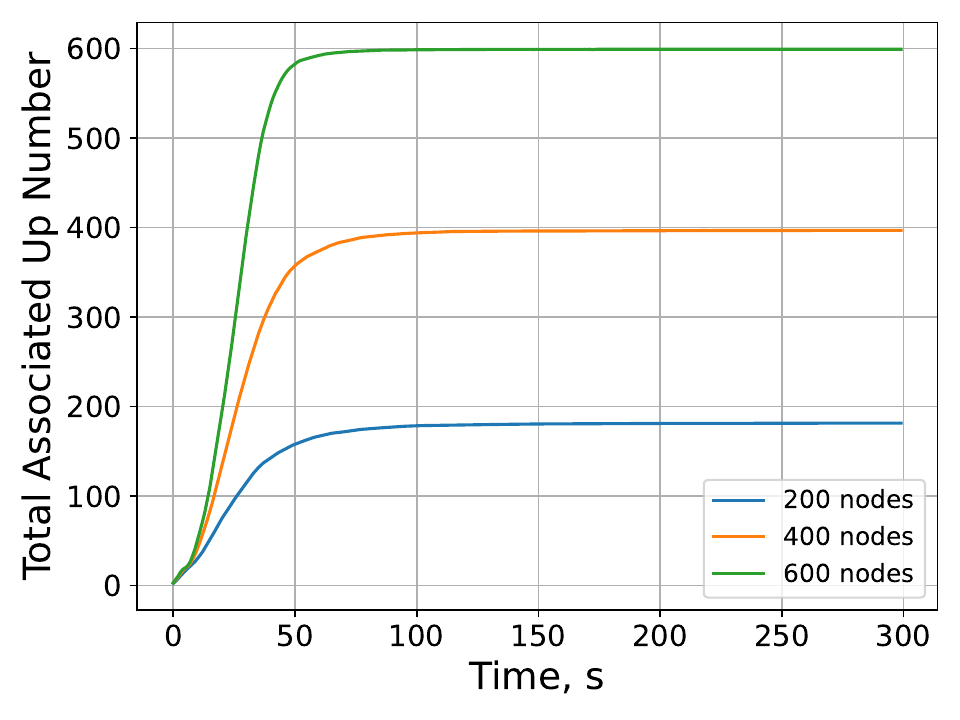}%
        \label{fig:totalRelay_inh_with}
    }
    \caption{Total number of associated nodes in the system for the mixed indoor-hall/outdoor propagation model.}
    \label{fig:totalRelay_inh}
    \vspace{-4mm}
\end{figure}

\section{Numerical Results}\label{sect:num}

In this section, we present our numerical results. We begin by evaluating the topology construction phase for both the normal and ultra-dense scenarios, considering cases with the proposed mechanisms both enabled and disabled. Specifically, we report the topology convergence time, association and reassociation rates, and structural parameters such as the number of active relays and the mean number of hops. 

Next, to assess the impact of this topology construction phase, we evaluate the data transmission phase metrics, including the PDR, mean packet delay, and the 95th percentile of packet delay, for both scenarios. Finally, we analyze the impact of enforcing energy-saving sleeping patterns at the RDs.% across both scenarios.

\subsection{Topology Construction Phase}

The proposed mechanisms can be divided into two groups. Load-dependent relay selection is aimed at ensuring connectivity for systems with different numbers of nodes, whereas hysteresis, blacklisting, and smart beaconing are specifically tailored for ultra-dense conditions. To this end, we first outline the impact of the former and then investigate the joint impact of all the mechanisms. 

\begin{figure}[!t]
    \centering
    \subfloat[All mechanisms disabled]{%
        \includegraphics[width=0.32\textwidth]{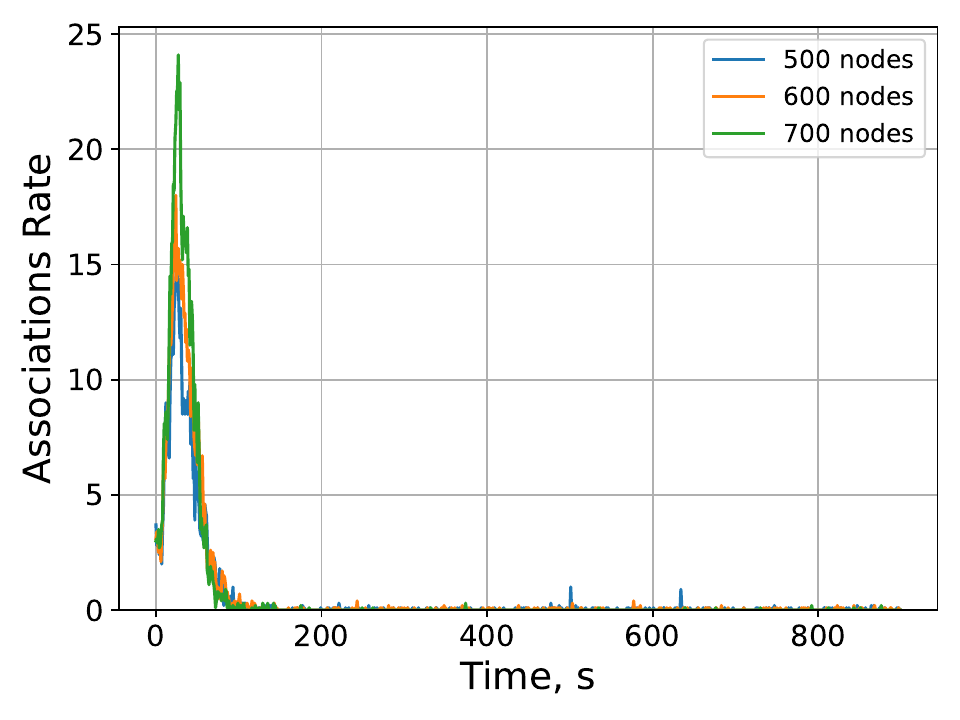}%
        \label{fig:ass_mixed_allDisabled}
    }\hfill
    \subfloat[All mechanisms enabled]{%
        \includegraphics[width=0.32\textwidth]{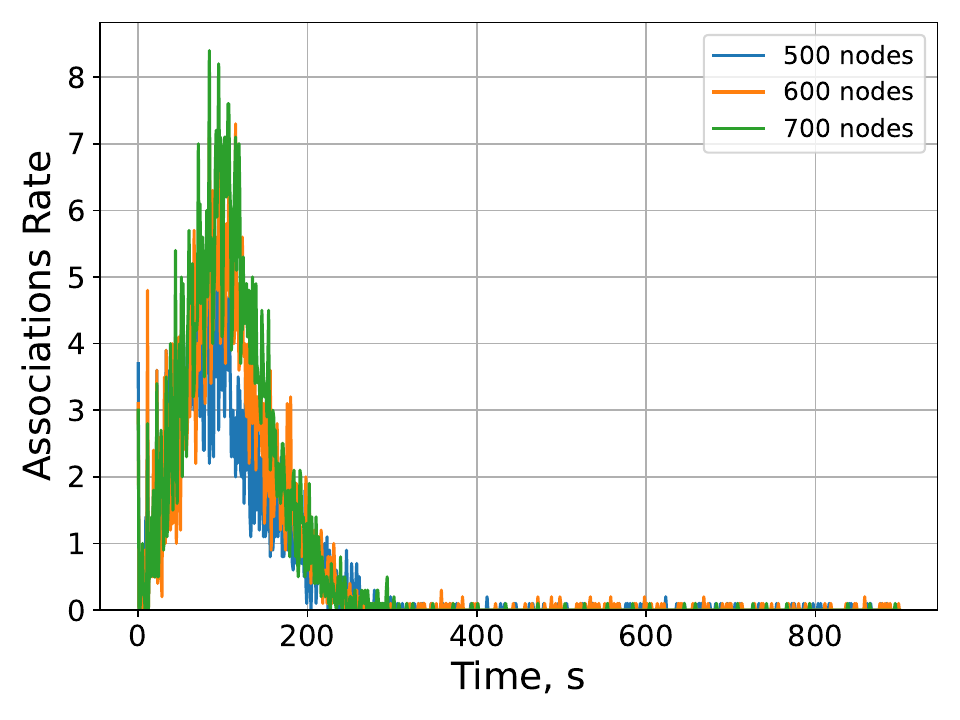}%
        \label{fig:ass_mixed_allEnabled}
    }
    \caption{Association rates for mixed indoor-hall/outdoor propagation model.}
    \label{fig:ass_mixed}
    \vspace{-0mm}
\end{figure}

\begin{figure}[!t]
    \centering
    \subfloat[All mechanisms disabled]{%
        \includegraphics[width=0.32\textwidth]{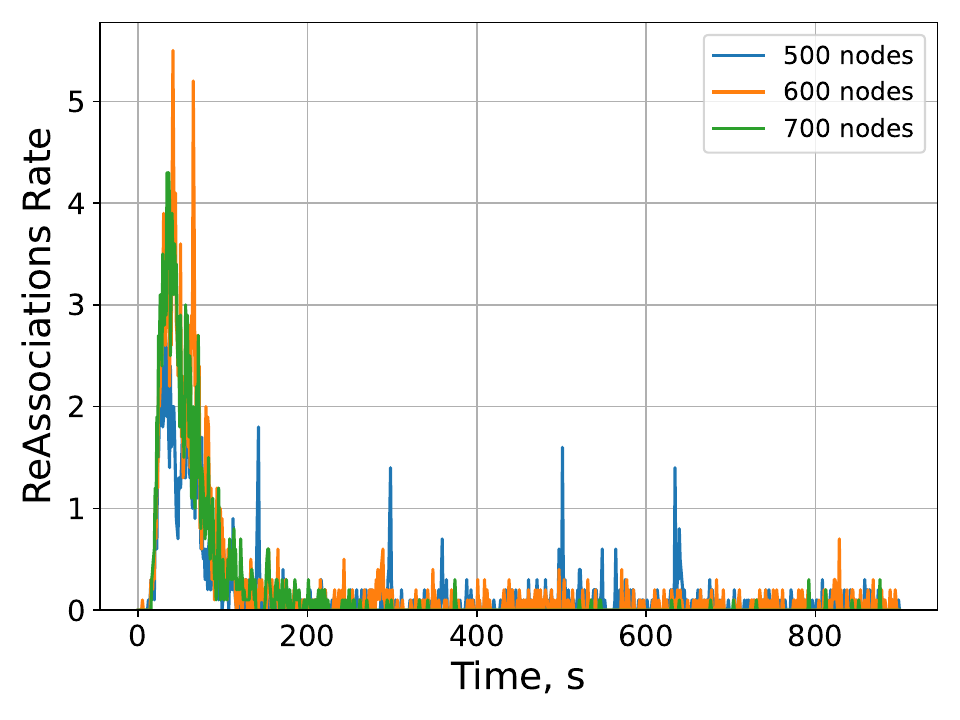}%
        \label{fig:reass_mixed_allDisabled}
    }\hfill
    \subfloat[All mechanisms enabled]{%
        \includegraphics[width=0.32\textwidth]{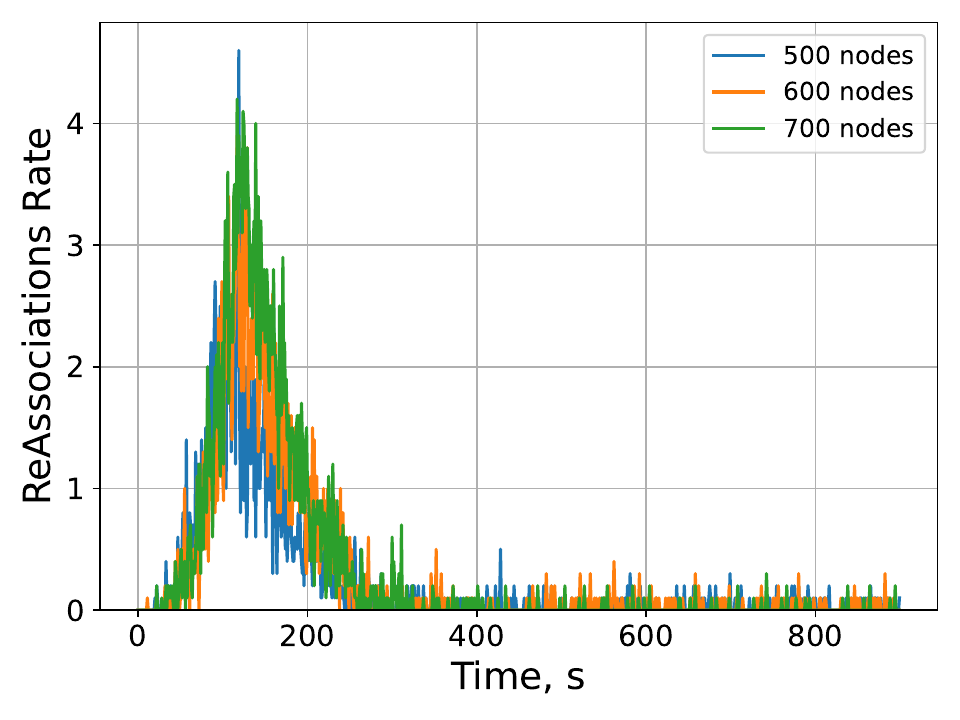}%
        \label{fig:reass_mixed_allEnabled}
    }
    \caption{Reassociation rates for mixed indoor-hall/outdoor propagation model.}
    \label{fig:reass_mixed_hall_outdoor}
    \vspace{-4mm}
\end{figure}

\begin{figure}[!t]
    \centering
    \subfloat[All mechanisms disabled]{%
        \includegraphics[width=0.32\textwidth]{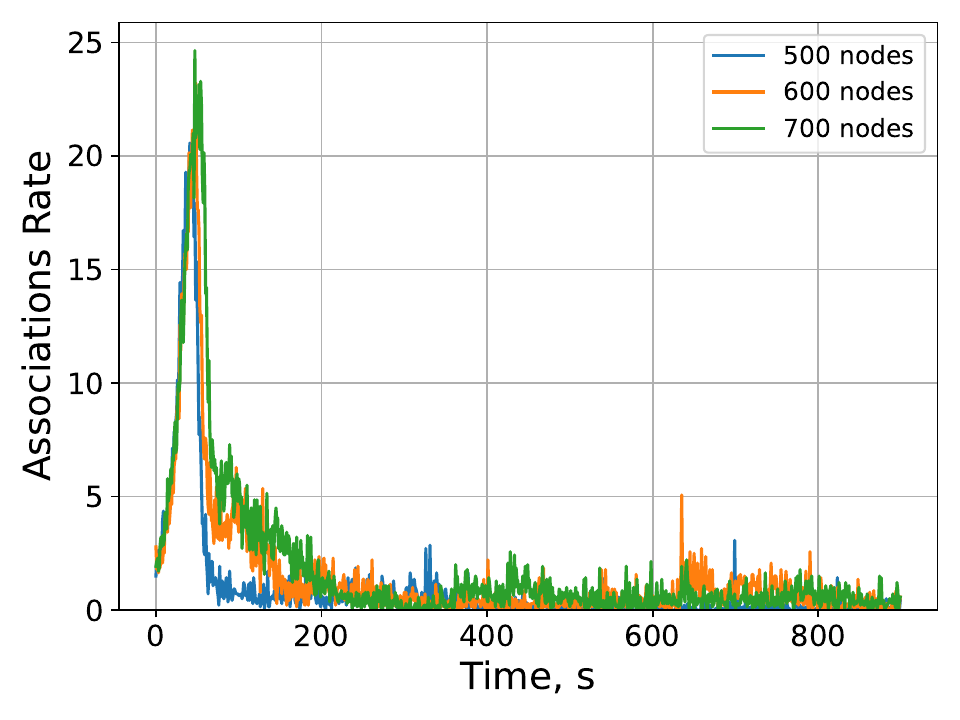}%
        \label{fig:ass_fspl_allDisabled}
    }\hfill
    \subfloat[All mechanisms enabled]{%
        \includegraphics[width=0.32\textwidth]{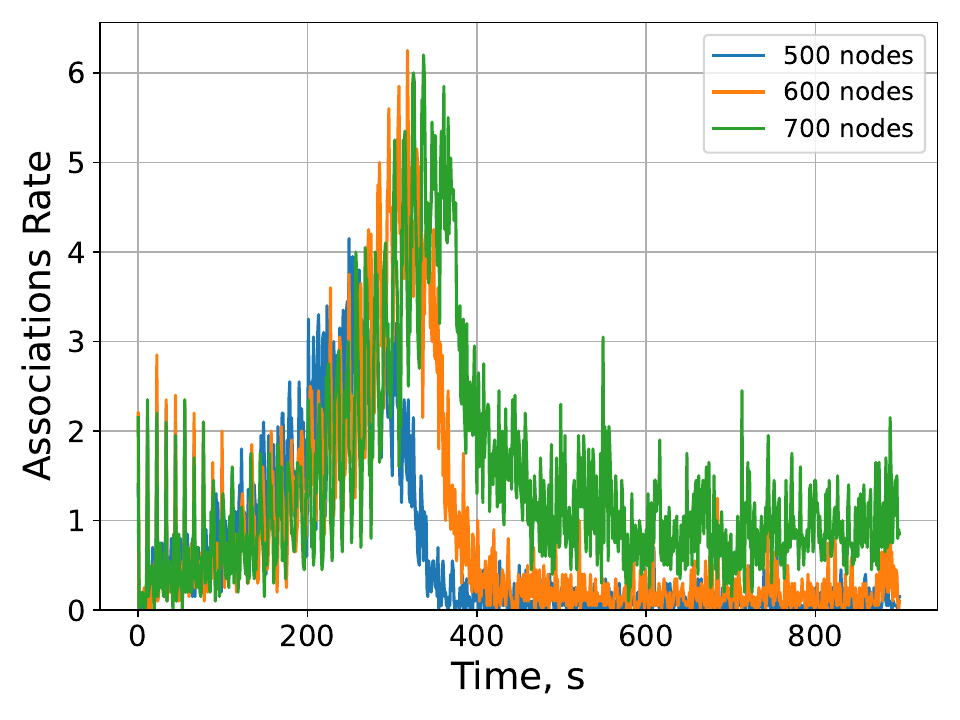}%
        \label{fig:ass_fspl_allEnabled}
    }
    \caption{Association rates for FSPL propagation model.}
    \label{fig:ass_fspl}
    \vspace{-4mm}
\end{figure}

\subsubsection{Load-Dependent Relay Selection}

% Mixed indoor-hall/outdoor

First, consider the proposed relay selection mechanism. To this end, Fig. \ref{fig:totalRelay_inh} shows the number of connected nodes for the mixed indoor-hall/outdoor propagation model, when the rest of the mechanisms are disabled. As one may observe, when load-dependent relay allocation is not utilized (Fig.~\ref{fig:totalRelay_inh_without}), not all nodes are connected. This could occur because of two effects: (i) an insufficient number of relays in the system and (ii) unfavorable geometric placement of the RDs. When load-dependent relay allocation is enabled, as shown in Fig. \ref{fig:totalRelay_inh_with}, all nodes in the 400- and 600-node systems become connected because relays are elected dynamically based on the local load. However, there is a small fraction of nodes, approximately 5--8\% in the 200-node deployment, that are still not connected. This is mainly caused by their geometrical locations, which do not allow a link to be established with nearby nodes. Thus, we may conclude that the proposed mechanism is critical for topology organization in DECT NR+ that allows dynamic adjustment to systems with different numbers of nodes. Owing to these observations, we will always keep this algorithm enabled in the following sections, as it addresses the principal behavior of the system. Note that for the FSPL propagation model, connectivity is not a principal issue because the entire coverage is a single interference domain, and thus, nodes may hear each other.

\begin{figure}[!t]
    \centering
    \subfloat[All mechanisms disabled]{%
        \includegraphics[width=0.32\textwidth]{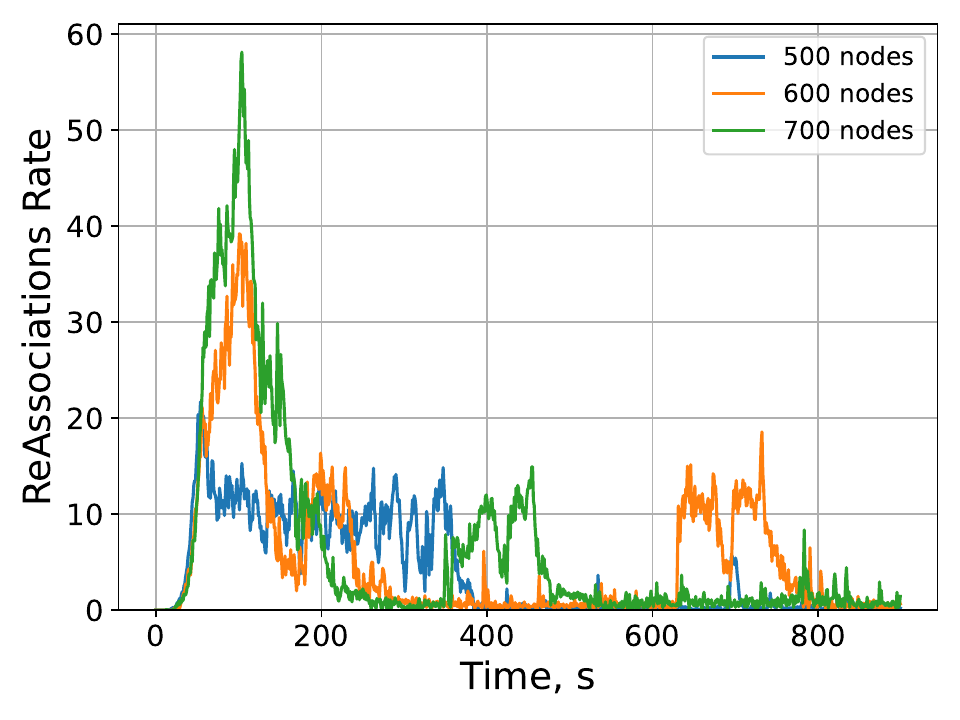}%
        \label{fig:reass_fspl_allDisabled}
    }\hfill
    \subfloat[All mechanisms enabled]{%
        \includegraphics[width=0.32\textwidth]{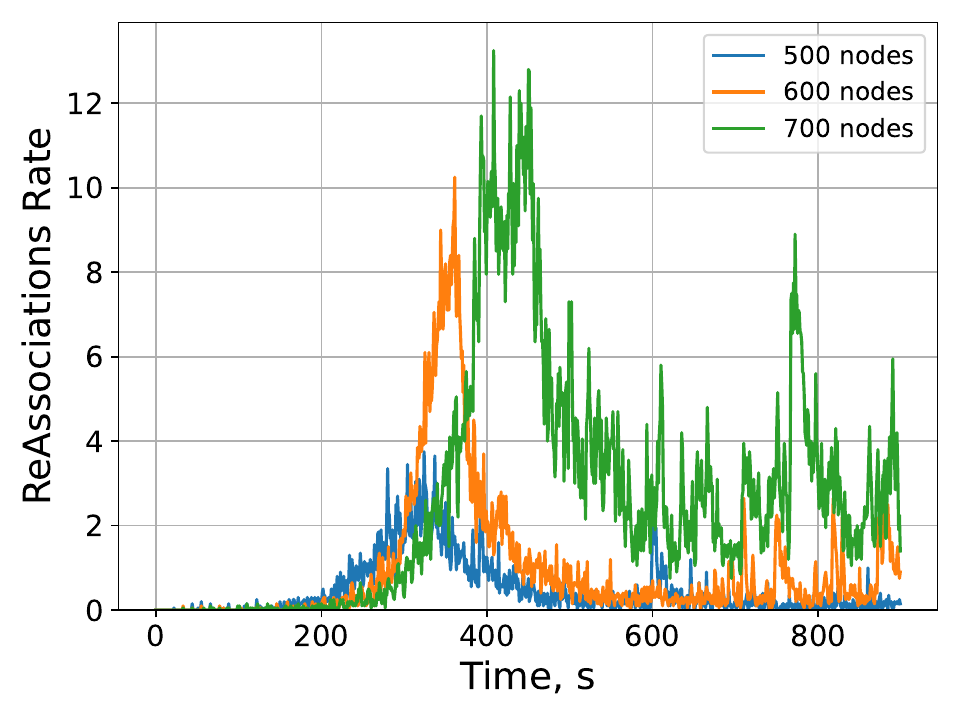}%
        \label{fig:reass_fspl_allEnabled}
    }
    \caption{Reassociation rates for FSPL propagation model.}
    \label{fig:reass_fspl}
    \vspace{-0mm}
\end{figure}

\begin{figure}[!t]
    \centering
    \subfloat[Mixed indoor-hall/urban model]{%
        \includegraphics[width=0.32\textwidth]{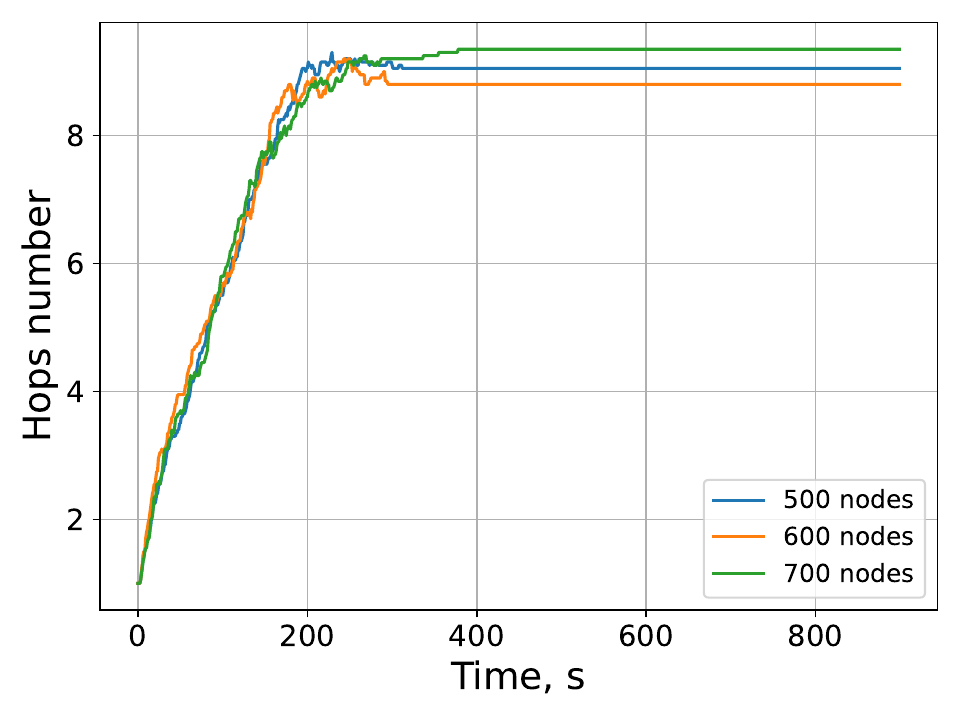}%
        \label{fig:NfsplHops}
    }\hfill
    \subfloat[FSPL model]{%
        \includegraphics[width=0.32\textwidth]{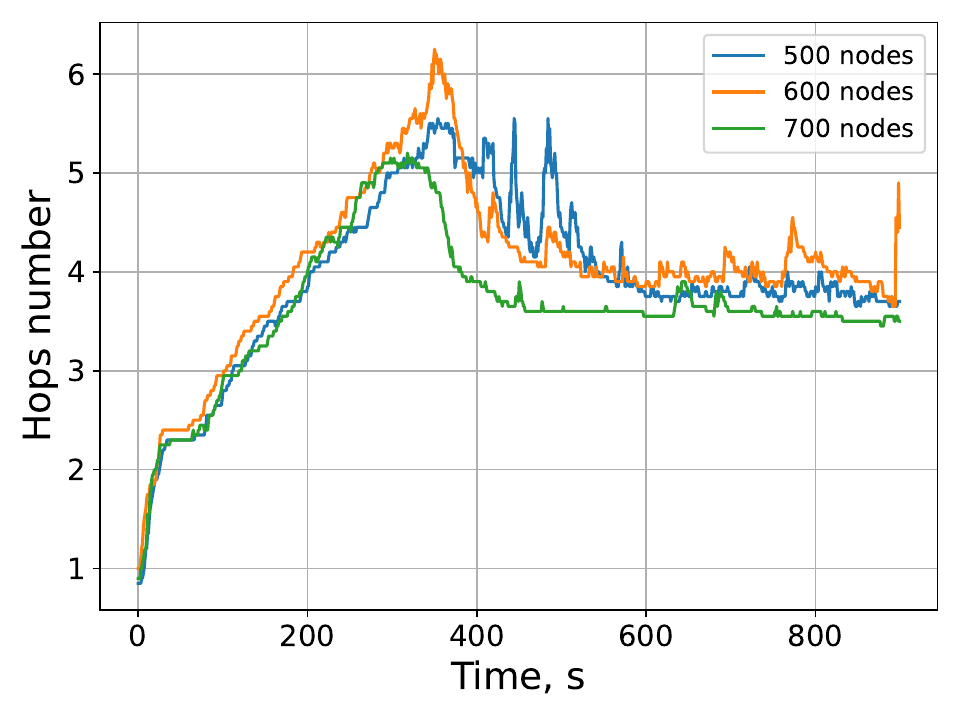}%
        \label{fig:fsplHops}
    }
    \caption{Number of hops for different propagation models.}
    \label{fig:fsplVsNfspl}
    \vspace{-4mm}
\end{figure}

\subsubsection{Normal Conditions: Mixed Indoor-Hall/Outdoor Model}

% One of the critical metrics for topology construction algorithm is association and reassociation rates, where the former shows how quickly topology is organized while the former specifies topology instability in time.

Now, consider the ultimate impact of the proposed mechanisms for the mixed indoor-hall/outdoor model. To this end, Fig. \ref{fig:ass_mixed} shows the association rates for two cases, when all the mechanisms (except for load-dependent relay allocation) are disabled and when they are all enabled. Recall that the greatest impact on the association rate should be produced by the blacklisting algorithm delaying reassociations in case of stale, unresponsive, or explicitly rejected association targets, and by smart beaconing allowing for the suppression of interference. 

By analyzing the presented results, it can be observed that in the absence of the proposed algorithms, the system completes the association phase approximately 200 s earlier. This is the impact of the blacklisting algorithm, which penalizes frequent reassociations by freezing nodes for a period of time. The peak association rate for 700 nodes in the system reaches 25 associations per second compared to approximately 8 associations per second when all mechanisms are enabled. Once all the nodes are associated, the association rate remains low for both schemes.

Consider now the reassociation rate for the mixed indoor-hall/outdoor model shown in Fig.~\ref{fig:reass_mixed_hall_outdoor} for two cases, when all mechanisms are disabled and enabled. We observe that in both cases, the peaks in the reassociation rates are co-located with the peaks in the association rates (see Fig.~\ref{fig:ass_mixed}). The impact of hysteresis is further observed in the tail of these illustrations. Specifically, when all mechanisms were enabled, large peaks in the reassociation rate were not present, as shown in Fig.~\ref{fig:reass_mixed_allEnabled}. As one may observe, even enabling all the mechanisms does not completely prevent network reconfiguration, still allowing it to occasionally find better paths to the sink.

\begin{figure}[!t]
    \centering
    \subfloat[Only load-dependent relay allocation enabled]{%
        \includegraphics[width=0.32\textwidth]{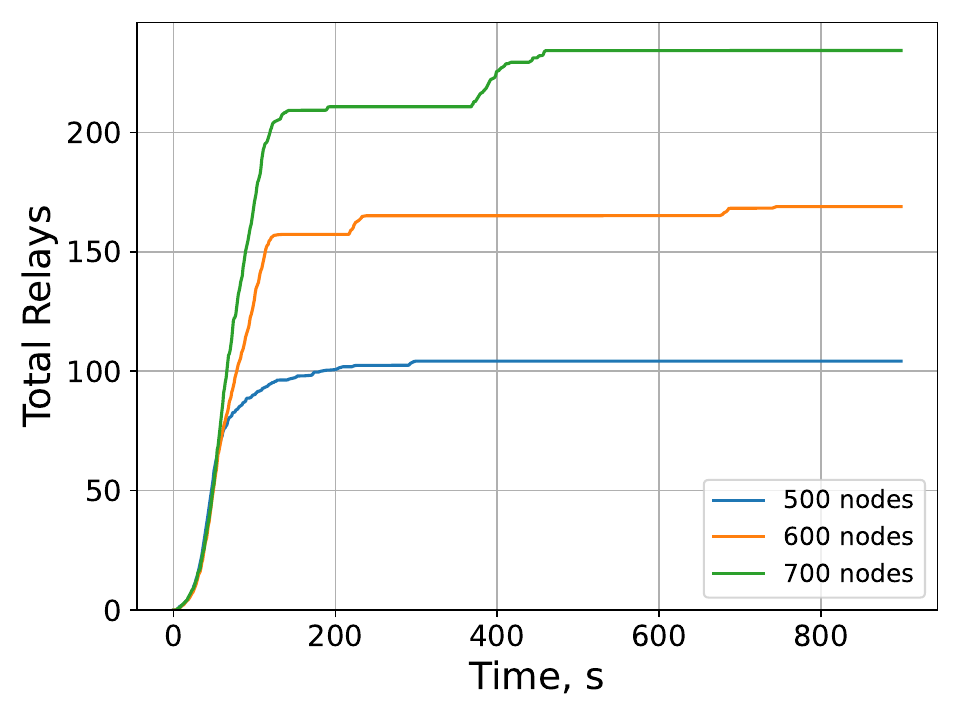}%
        \label{fig:relays_fspl_allDisabled}
    }\hfill
    \subfloat[All mechanisms enabled]{%
        \includegraphics[width=0.32\textwidth]{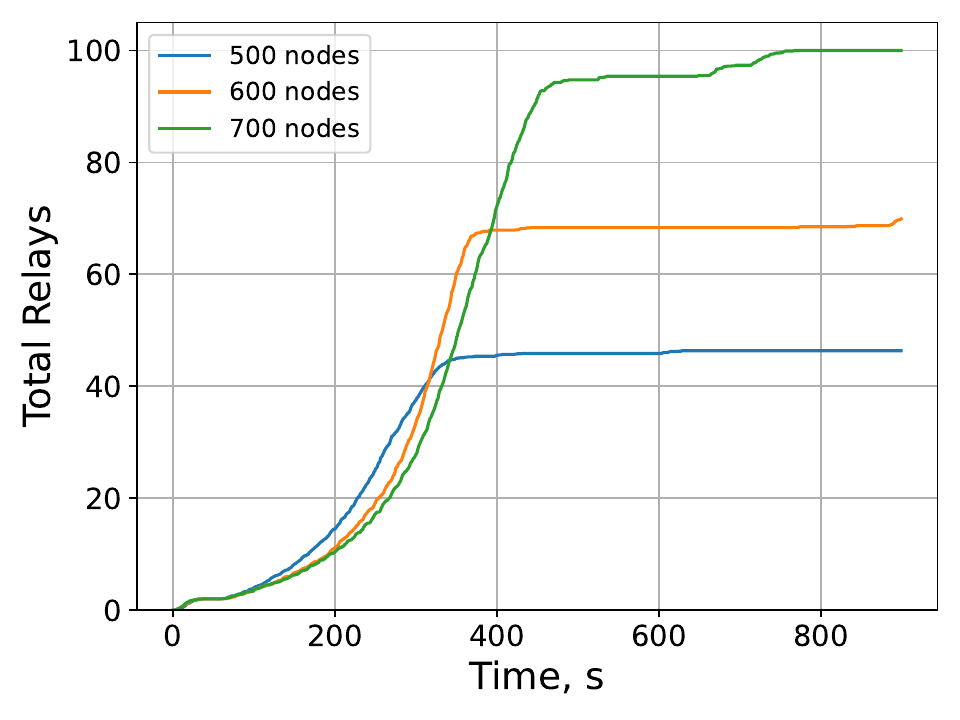}%
        \label{fig:relays_fspl_allEnabled}
    }
    \caption{Number of relays in the system for FSPL propagation model.}
    \label{fig:relays_fspl}
    \vspace{-4mm}
\end{figure}

\begin{figure*}[!t]
    \centering
    \subfloat[Packet delivery ratio]{%
        \includegraphics[width=0.3\textwidth]{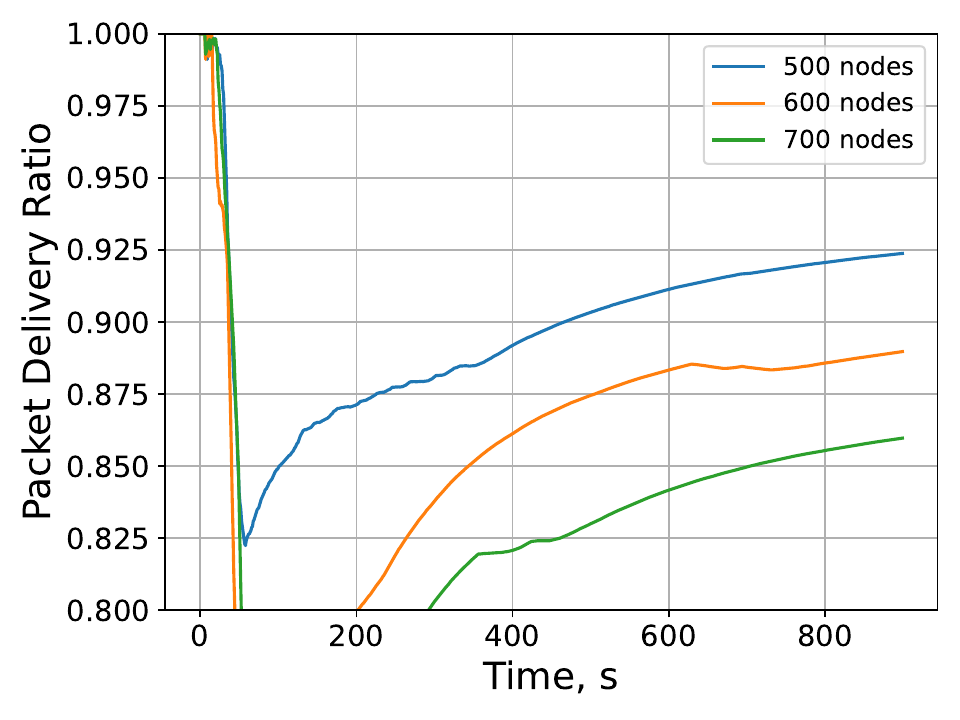}%
        \label{fig:data_fspl_allDisabled_pdr}
    }\hfill
    \subfloat[Mean packet delay]{%
        \includegraphics[width=0.3\textwidth]{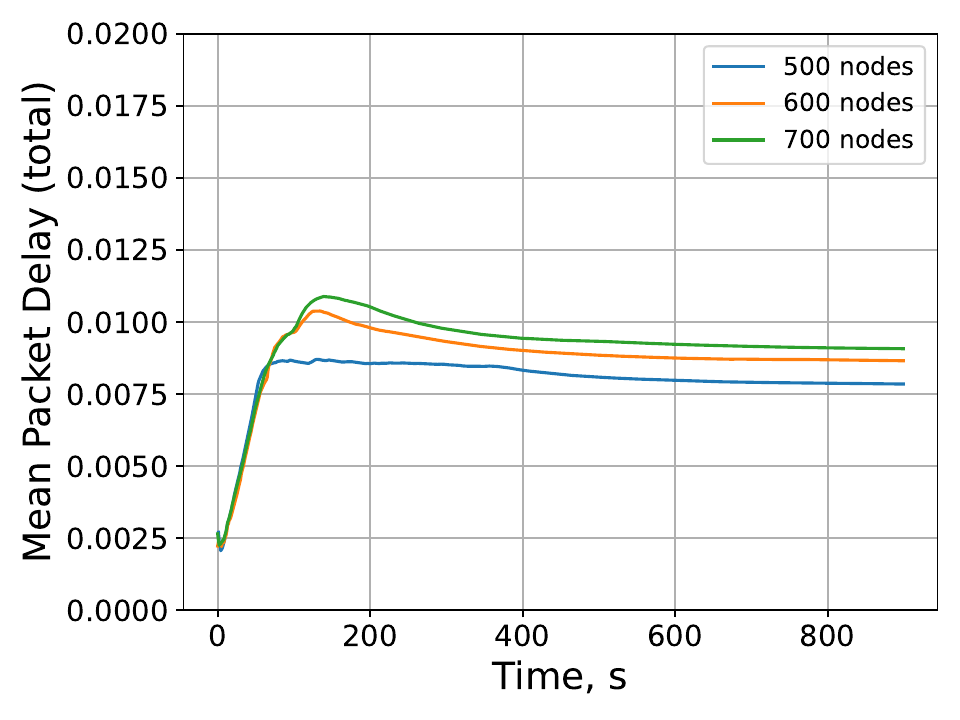}%
        \label{fig:data_fspl_allDisabled_mean}
    }\hfill
    \subfloat[95th percentile packet delay]{%
        \includegraphics[width=0.3\textwidth]{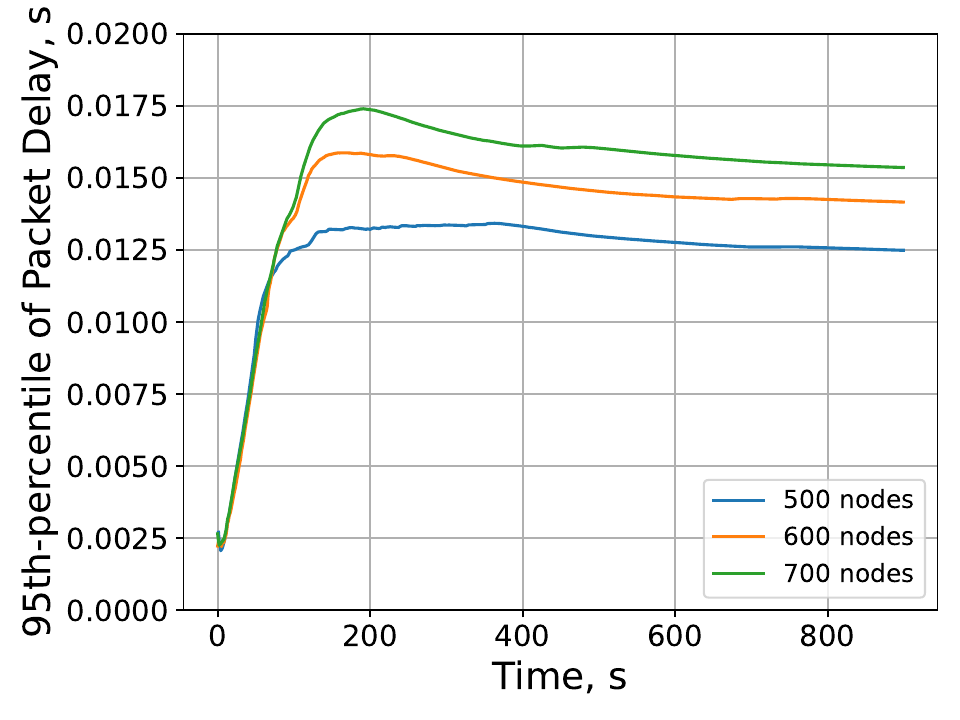}%
        \label{fig:data_fspl_allDisabled_quantile}
    }
    \caption{Data transmission performance for FSPL propagation model with all mechanisms disabled.}
    \label{fig:data_fspl_allDisabled}
    \vspace{-2mm}
\end{figure*}

\begin{figure*}[!t]
    \centering
    \subfloat[Packet delivery ratio]{%
        \includegraphics[width=0.3\textwidth]{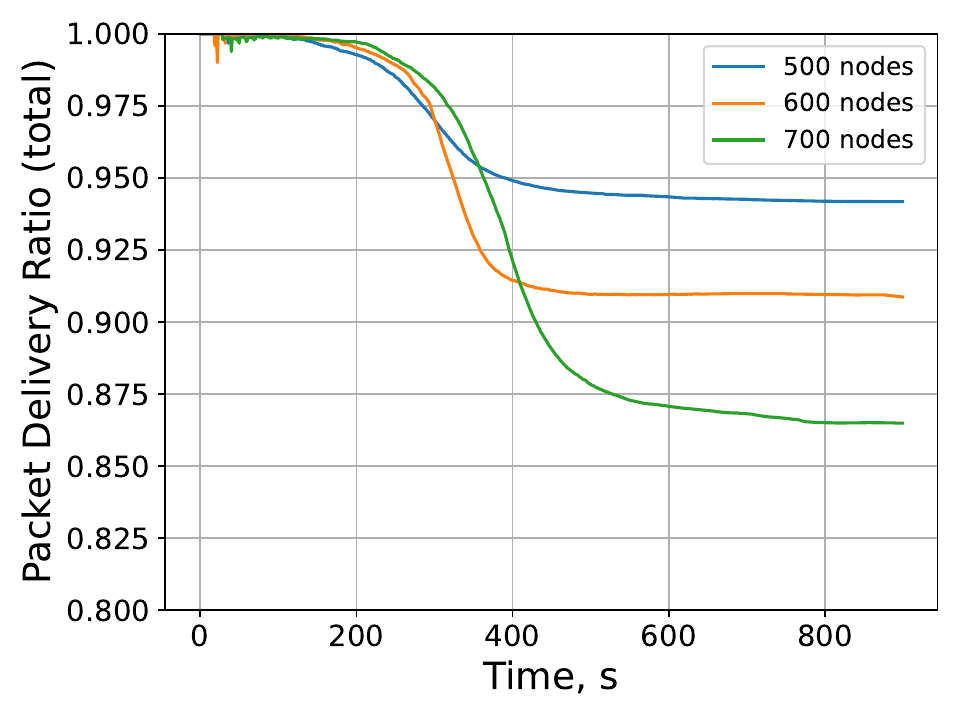}%
        \label{fig:data_fspl_allEnabled_pdr}
    }\hfill
    \subfloat[Mean packet delay]{%
        \includegraphics[width=0.3\textwidth]{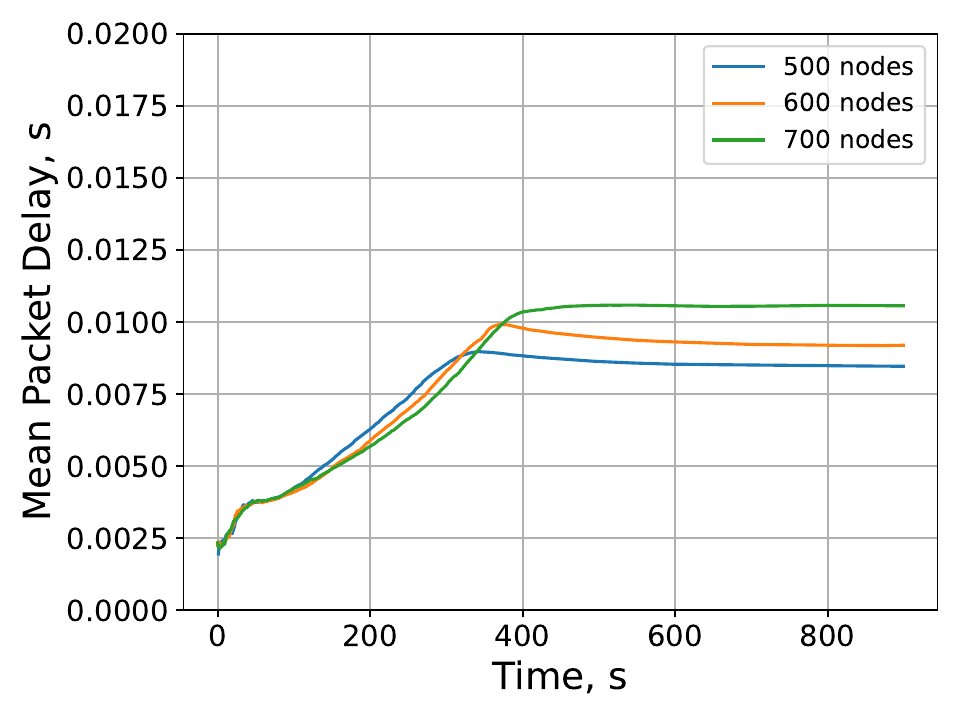}%
        \label{fig:data_fspl_allEnabled_mean}
    }\hfill
    \subfloat[95th percentile packet delay]{%
        \includegraphics[width=0.3\textwidth]{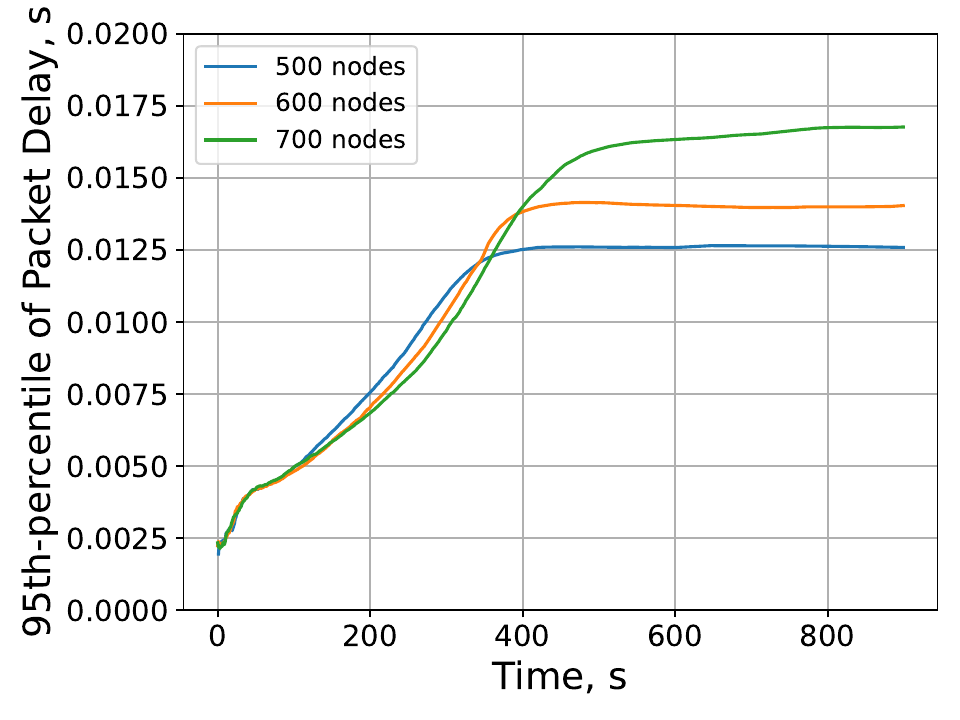}%
        \label{fig:data_fspl_allEnabled_quantile}
    }
    \caption{Data transmission performance for FSPL propagation model with all mechanisms enabled.}
    \label{fig:data_fspl_allEnabled}
    \vspace{-4mm}
\end{figure*}

\subsubsection{Ultra-Dense Conditions: FSPL Model}

Consider now the association rate for the ultra-dense scenario caused by FSPL propagation, as shown in Fig. \ref{fig:ass_fspl}. Here, we observe that disabling all the mechanisms, similar to the mixed indoor-hall/outdoor propagation model, leads to a higher peak in the association rate. However, the behavior of the metric is no longer smooth after this point. Specifically, we see that the association rate, even for 500 and 600 nodes in the system, fluctuates, leading to association rates higher than 1--2 associations per second. This behavior is observed only for 700 nodes when all mechanisms are enabled, and even in this case it varies around just 1 association per second.

To obtain a clearer picture of the topology, we consider Fig. \ref{fig:reass_fspl}, which shows the reassociation rates for the considered cases under the FSPL model. Fig.~\ref{fig:reass_fspl_allDisabled} shows that the topology might be highly unstable when the proposed mechanisms are not utilized. Specifically, the reassociation rate may occasionally exceed 10 reassociations per second, even for systems with 500 and 600 nodes, causing drastic topology reconfigurations. This may lead to instability of the data transmission phase and may result in increased power consumption by nodes. By applying the proposed mechanisms, one may stabilize the topology such that even for 700 nodes in the system, the reassociation rate varies around 2--3 reassociations per second, while for a smaller number of nodes, it stays below 1 reassociation per second.

\subsubsection{Number of Hops}

% FAPL vs non-FSPL

One critical measure that allows the assessment of the quality of the topology is the number of hops to the sink. For a mixed indoor-hall/outdoor propagation model, this quantity should naturally be larger than that for the FSPL model, since worse propagation forces longer multi-hop routes. We now check it in Fig. \ref{fig:fsplVsNfspl} for the case when all considered mechanisms are enabled. The numbers of nodes in the compared cases are set to 500, 600, and 700. We observe that for the mixed indoor-hall/urban propagation model, that is, for non-dense deployment conditions, the number of hops remains at approximately 9 for all the considered numbers of nodes, as shown in Fig. \ref{fig:NfsplHops}. However, for the FSPL model, the mean number of hops decreases to just four. Furthermore, as shown in Fig.~\ref{fig:fsplHops}, the system does not settle on the first constructed topology. The beacon placement and association and relay selection algorithms continue to work together, converging to shorter routes.

% We conclude our discussion of topology-related metrics by cross-comparing the mean number of hops in the system with the FSPL propagation model and the system with the mixed indoor-hall/urban propagation model, with both systems utilizing the proposed approach. The rationale for this comparison is that the FSPL should naturally lead to a smaller number of hops in the system.

\subsubsection{Number of Relays}

Note that the use of load-dependent relay allocation may also lead to an unwanted increase in the number of relays in the system for ultra-dense topologies. This is shown in Fig. \ref{fig:relays_fspl_allDisabled} illustrating the number of relays for a DECT NR+ system with different numbers of nodes when all the proposed algorithms except for load-dependent relay allocation are disabled. As one may observe, the fraction of nodes acting as relays is extremely large, reaching approximately 34\% for a system with 700 nodes. This could eventually lead to excessive power consumption of the entire system and reduce the network lifetime. However, as shown in Fig. \ref{fig:relays_fspl_allEnabled}, this behavior is effectively suppressed by enabling the rest of the algorithms. Specifically, for a system with 700 nodes, the number of nodes elected as relays decreases from around 240 to 100, that is, approximately 15\% of all nodes. For a system with 500 nodes, the fraction of relays is 10\%.

\subsection{Data Transmission Phase}

Topology-related metrics alone cannot provide a complete characterization of the mesh system performance. To this end, we investigate the performance parameters delivered by the proposed algorithms at the data transmission phase.

We begin with the FSPL propagation model and consider the case when all mechanisms are disabled, as shown in Fig. \ref{fig:data_fspl_allDisabled}. Here, we observe a sharp drop in the PDR when the system enters operation. By comparing the timing of this drop with the data shown in Fig. \ref{fig:ass_fspl_allDisabled}, we deduce that it is caused by nodes that start generating traffic immediately after establishing a route to the sink. The root cause is that data packets start colliding with both beacons and association request packets. Although the system eventually recovers, it takes more than 10 min for the PDR to plateau. Similar spikes are also evident in the mean delay and the 95th percentile of packet delay results shown in Figs. \ref{fig:data_fspl_allDisabled_mean} and \ref{fig:data_fspl_allDisabled_quantile}.

Now, consider what happens when we turn on the proposed mechanisms. The results for the FSPL propagation model are shown in Fig. \ref{fig:data_fspl_allEnabled}. As shown in Fig.~\ref{fig:data_fspl_allEnabled_pdr}, the PDR now decreases gradually to the steady-state level without a drastic drop at the beginning. This is mainly because both the blacklisting mechanism delays the reassociations for nodes and the smart beacon positioning reduces excessive interference at the association phase. Note that in absolute numbers, there is a slight advantage compared to the case when all mechanisms are turned off. However, this is negligible for the considered load of 10 packets per second. The same smooth behavior without jumps is observed for the mean latency and the 95th percentile of latency. Note that the observed absolute values are slightly better for the system with all mechanisms enabled. %However, this occurs because of the lower PDR.

% \textcolor{red}{RG: reconcile with Table~\ref{tab:simScen} traffic model (bursts of 2, mean interpacket 10--50 s). State whether 10 pkt/s is per node or aggregate.}

\begin{figure*}[!t]
    \centering
    \subfloat[Packet delivery ratio]{%
        \includegraphics[width=0.3\textwidth]{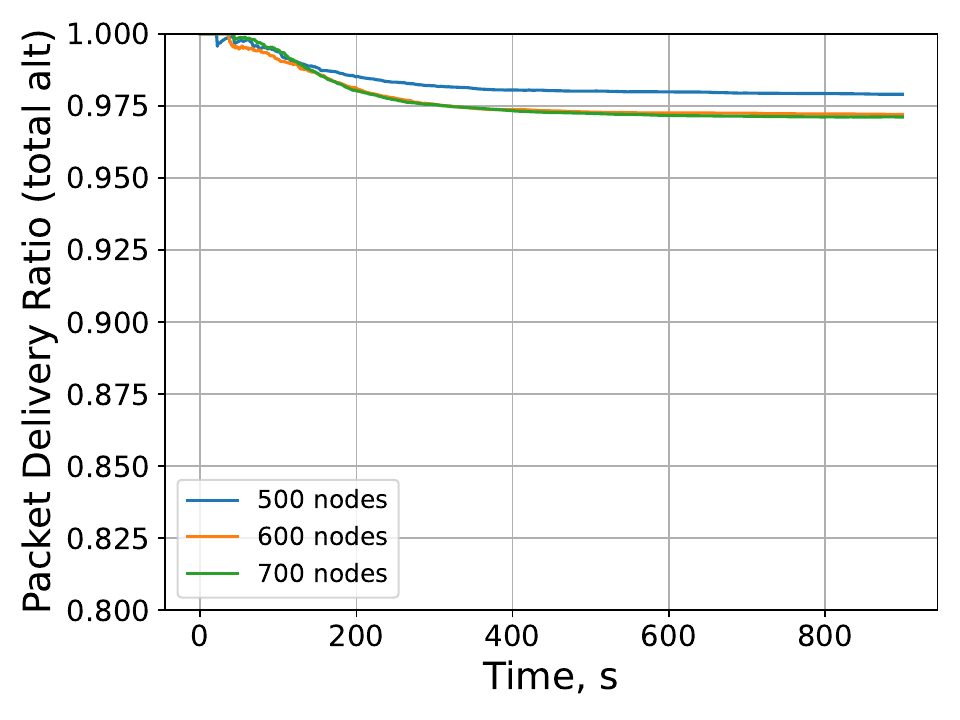}%
        \label{fig:data_mixed_allEnabled_pdr}
    }\hfill
    \subfloat[Mean packet delay]{%
        \includegraphics[width=0.3\textwidth]{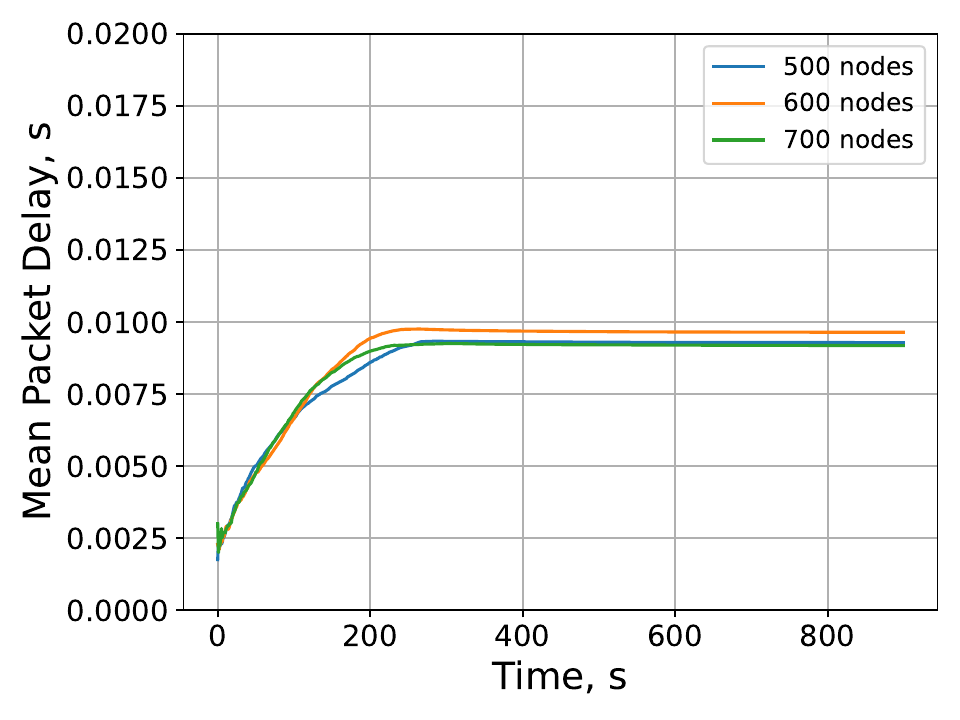}%
        \label{fig:data_mixed_allEnabled_mean}
    }\hfill
    \subfloat[95th percentile packet delay]{%
        \includegraphics[width=0.3\textwidth]{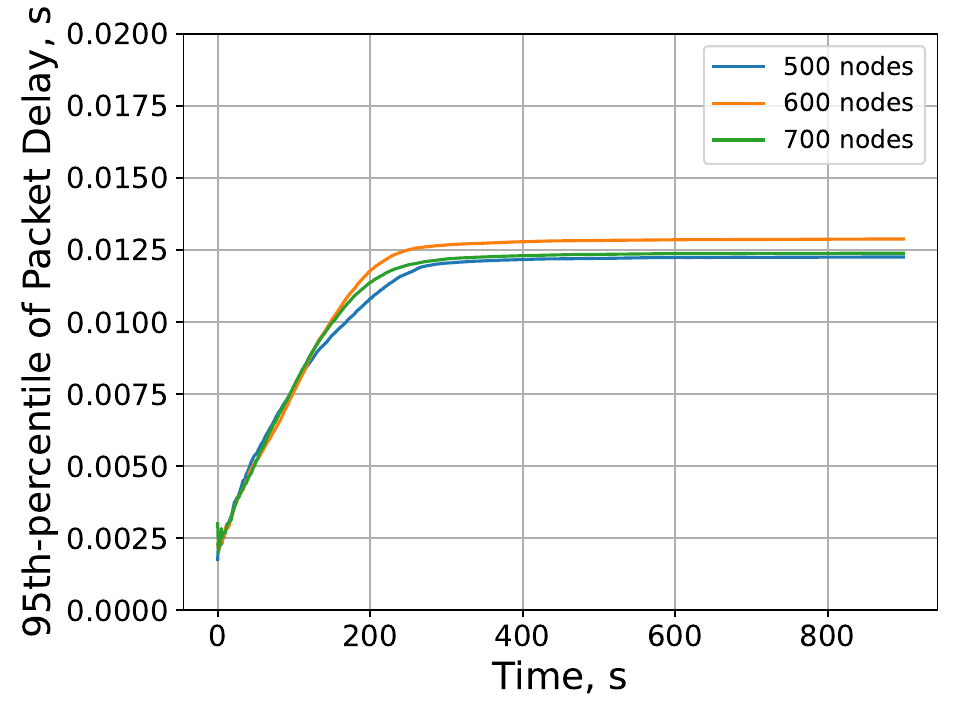}%
        \label{fig:data_mixed_allEnabled_quantile}
    }
    \caption{Data transmission performance for mixed indoor-hall/outdoor propagation model with all mechanisms enabled.}
    \label{fig:data_mixed_allEnabled}
    \vspace{-2mm}
\end{figure*}

As a benchmark, we compared the reported results for the FSPL propagation model with those of the mixed indoor-hall/outdoor model, as shown in Fig. \ref{fig:data_mixed_allEnabled}. By comparing the results shown in Figs. \ref{fig:data_fspl_allEnabled} and \ref{fig:data_mixed_allEnabled}, we note that FSPL does lead to worse results in terms of both PDR and latency measures. The degradation in the PDR for the FSPL model is proportional to the number of nodes in the system, implying that the main reason is the increased interference owing to significantly better propagation conditions. In contrast, there are no significant differences between the PDRs for the mixed indoor-hall/outdoor model, as the nodes are much better isolated from each other in terms of interference.

It is interesting to note that the mean latency and the 95th percentile of latency are similar. However, these latencies originate from different dominant sources. Recalling the results shown in Fig. \ref{fig:fsplVsNfspl}, we see that the major component in the case of the mixed indoor-hall/outdoor propagation model is the number of hops to the sink. In contrast, for the FSPL model, the number of hops is only four on average, implying that a significant part of the latency comes from link retransmissions caused by interference.

\begin{figure*}[!t]
    \centering
    \subfloat[Association rate]{%
        \includegraphics[width=0.3\textwidth]{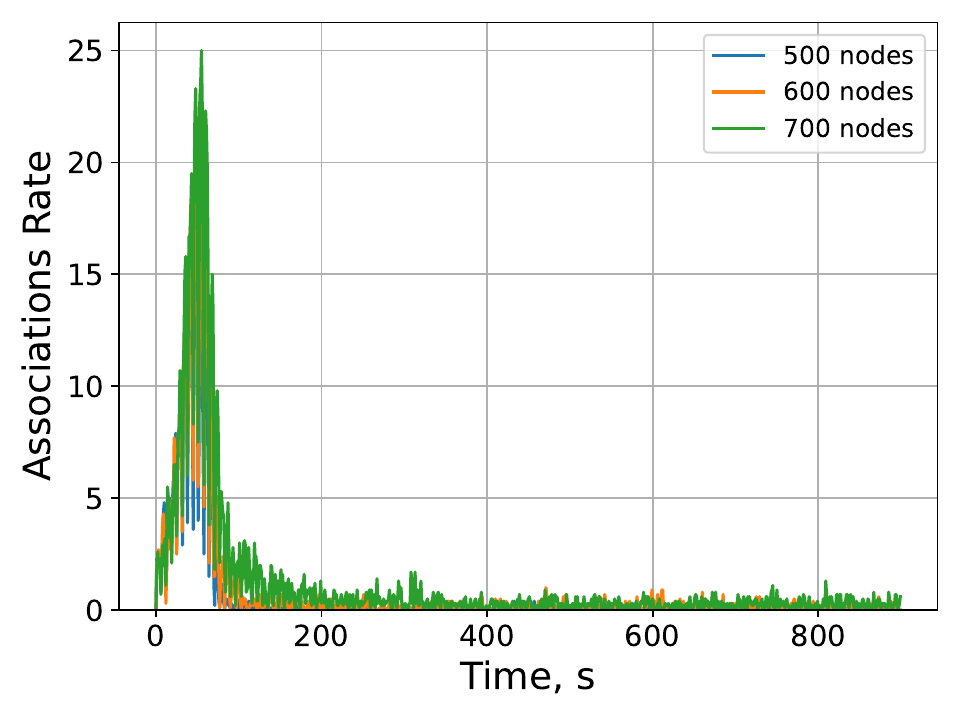}%
        \label{fig:data_assoc_sleeping}
    }\hfill
    \subfloat[Packet delivery ratio]{%
        \includegraphics[width=0.3\textwidth]{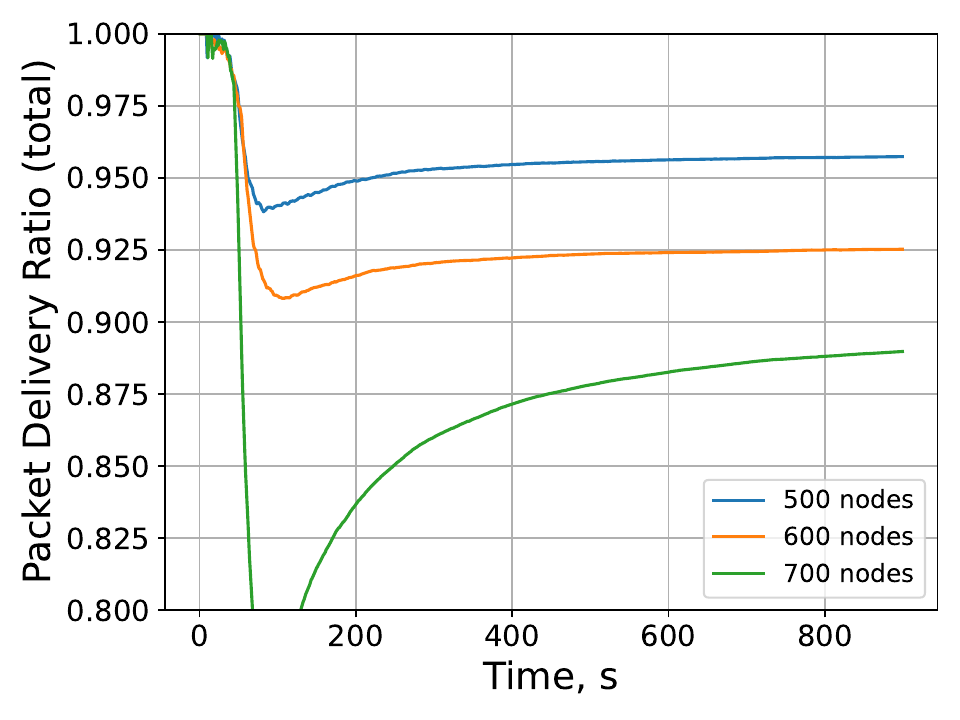}%
        \label{fig:data_pdr_sleeping}
    }\hfill
    \subfloat[95th percentile packet delay]{%
        \includegraphics[width=0.3\textwidth]{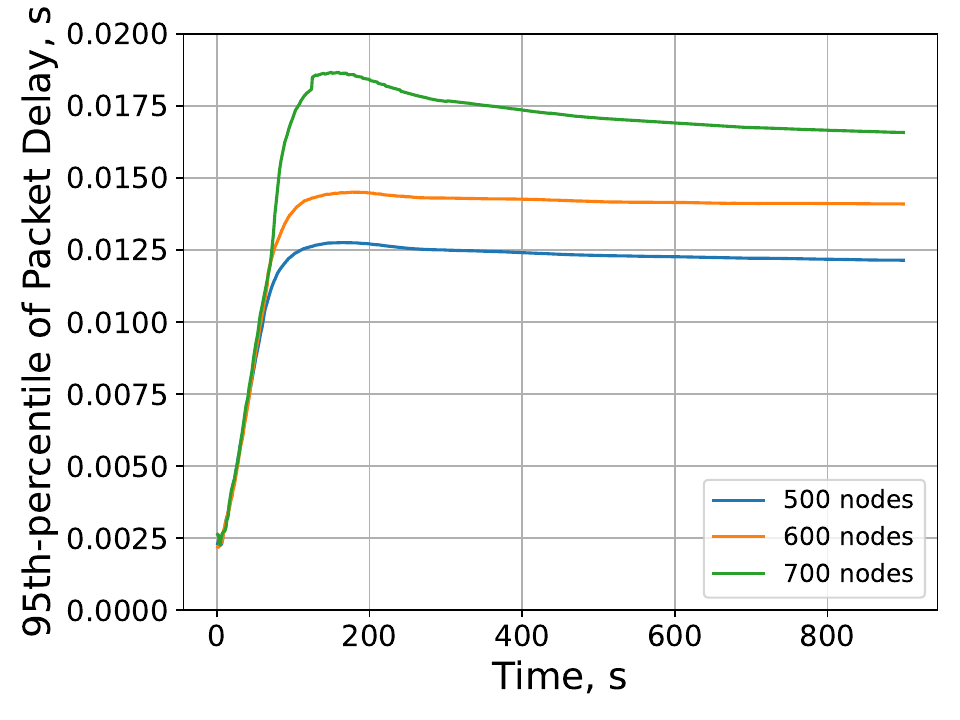}%
        \label{fig:data_delay_quant_sleeping}
    }
    \caption{Topology- and data-transmission-related metrics for a system with periodically sleeping leaf RDs.}
    \label{fig:data_sleeping}
    \vspace{-4mm}
\end{figure*}

\subsection{System with Periodically Sleeping Nodes}

DECT NR+ can be used in deployments where nodes operate on batteries. In this case, the leaf RDs may utilize sleep patterns. At this final stage, we evaluate the system behavior under periodic sleeping patterns, where all RDs asynchronously alternate between 5 s of sleep and 1.5 s of active operation, resulting in an active-time fraction of approximately 23\%. Note that the active period duration is chosen such that upon waking up, the node will detect a beacon from the relay with which it is currently associated. The beaconing interval is set to 1 s.

Fig. \ref{fig:data_sleeping} shows the main topology- and data-transmission-related metrics, including association rate, PDR, and the 95th percentile of packet delay for the system with all proposed mechanisms enabled under the FSPL propagation model. First, as shown in Fig. \ref{fig:data_assoc_sleeping}, the association performance drastically improves. Specifically, by cross-comparing with the data shown in Fig. \ref{fig:ass_fspl_allEnabled}, we see that there is no delay caused by the blacklisting algorithm. The rationale is that the sleeping patterns on RDs lead to much smaller interference in the system, as all RDs follow sleep patterns asynchronously. This behavior leads to fewer nodes entering the blacklisting mode and eventually results in all the nodes being connected by approximately 200 s for the system with 700 nodes.

The described gains also affect the performance in the data transmission phase. By cross-comparing the results shown in Fig. \ref{fig:data_pdr_sleeping} to those demonstrated in Figs. \ref{fig:data_fspl_allDisabled_pdr} and \ref{fig:data_fspl_allEnabled_pdr} for the same propagation model, we see that the system behavior begins to resemble the case in which all the proposed mechanisms are disabled, but the time required for the PDR to reach steady state is approximately halved as compared even to the case when all the mechanisms are activated. Specifically, the convergence time is now approximately 200 s, as compared to Fig. \ref{fig:data_fspl_allEnabled_pdr}, where it is approximately twice as long, around 400 s. The root cause is periodic sleeping, leading to milder interference conditions during the warm-up time. Similar observations can be made for the 95th percentile of packet delay. Finally, we note that the absolute values of the PDR and delay quantities in the steady-state regime are similar to those of the system without periodic sleeping.

% ADDED CONCLUSION
% - the system, where RDs undergo period sleeping pattern is characterized with approximately twice better performance in terms of convergence of topology and data transmission metrics.
% - 
% - 

\section{Conclusions}

This study addresses the inherent instability challenges in 5G-grade DECT NR+ mesh networks under ultra-dense deployments enabled by exceptional propagation conditions. Conventional decentralized topology construction algorithms are highly susceptible to dense node environments, where interference during the association phase triggers topology reconfigurations. By introducing a joint suite of mechanisms, including load-dependent relay allocation, hysteresis- and blacklist-based association control, and smart beacon placement, this study successfully stabilizes the network topology. System-level simulations demonstrate that these mechanisms mitigate the interference caused by chaotic association dynamics and beacon overlap, ensuring efficient network scaling.

The numerical evaluation highlights the profound impact of the proposed algorithms on the structural network efficiency. In unoptimized dense networks, the reassociation rate can surge beyond 10 reassociations per second for deployments of 500 to 600 nodes, leading to volatile topological shifts. Implementing the proposed mechanisms suppresses this volatility, maintaining a stable reassociation rate of approximately 2--3 per second even in the 700-node scenario and keeping it well below 1 reassociation per second for smaller networks. Furthermore, the load-dependent relay allocation dynamically optimizes network resources, effectively reducing the required number of active relays by more than half, from 34\% to 15\%.

Ultimately, stabilizing the topology yields performance gains during the subsequent data transmission phase. Without these optimizations, association dynamics cause drops in the PDR and substantial latency spikes, from which the system recovers slowly, taking over 10 min to reach the steady state. The proposed joint mechanisms successfully avert these performance degradations, leaving only the inherent interference caused by the actual data delivery. Consequently, DECT NR+ can reliably support mMTC services across large-scale networks of several hundred nodes, even in challenging environments such as airborne systems.

%The important application areas of the recently standardized DECT NR+ systems are operation in extremely good propagation conditions (e.g., intra-UAV swarm communications) or in extremely dense deployments. However, the baseline topology formation algorithms specified in the DECT NR+ standards are not optimized to deliver a potentially feasible quality of service to the end nodes. The main prohibiting factor is extremely high interference at the node association phase, resulting in sporadic fluctuations in the topology.

%To address this issue we proposed a new beacon transmission time selection scheme that allows to ensure that most of those transmitted by relays are not overlapped. With the proposed approach, ultradense DECT NR+ systems can deliver quality of service in terms of PDF and delay metrics comparable to those of systems with either a limited number of nodes or significantly worse propagation conditions. However, under these conditions, the system becomes much more sensitive to the number of nodes.

%\section*{Acknowledgment}

%This work was supported by the Business Finland project "AURORA.”

%\balance
\bibliographystyle{IEEEtran}
\bibliography{reference}

\begin{IEEEbiography}[{\includegraphics[width=1in,height=1.25in,clip,keepaspectratio]{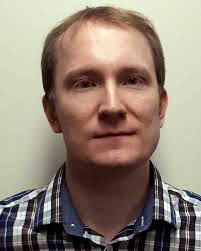}}]{Andrey Samuylov} received his MSc in Applied Mathematics and Cand.Sc. in Physics and Mathematics from the RUDN University, Russia, in 2012 and 2015, respectively. Since 2015 he is working at Tampere University of Technology as a researcher, working on analytical performance analysis of various 5G wireless networks technologies. His research interests include P2P networks performance analysis, performance evaluation of wireless networks with enabled D2D communications, and mmWave-band communications.
\end{IEEEbiography}

\biographyspace

\begin{IEEEbiography}[{\includegraphics[width=1in,height=1.25in,clip,keepaspectratio]{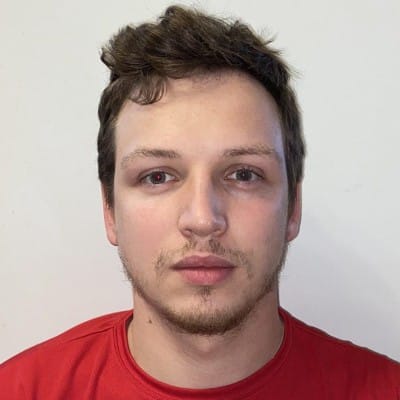}}]{Roman Glazkov} received the MSc degree in Radio Communications, Radio Broadcasting and Television from the St. Petersburg State University of Telecommunications, Russia, in 2013. Currently, he is a Doctoral Researcher with the Faculty of Information Technology and Communication Sciences, Tampere University, Finland. Previously, he held research, technical management, and network architecture positions with YL-Verkot Oy and Huawei. He has (co)authored publications on wireless communications, 5G and beyond networks, edge and fog computing, software-defined radio, and DECT-2020 NR systems. His research interests include DECT-2020 NR, near-URLLC services, multi-hop wireless systems, 5G/6G networks, agentic AI, and edge computing.
\end{IEEEbiography}

\biographyspace

\begin{IEEEbiography}[{\includegraphics[width=1in,height=1.25in,clip,keepaspectratio]{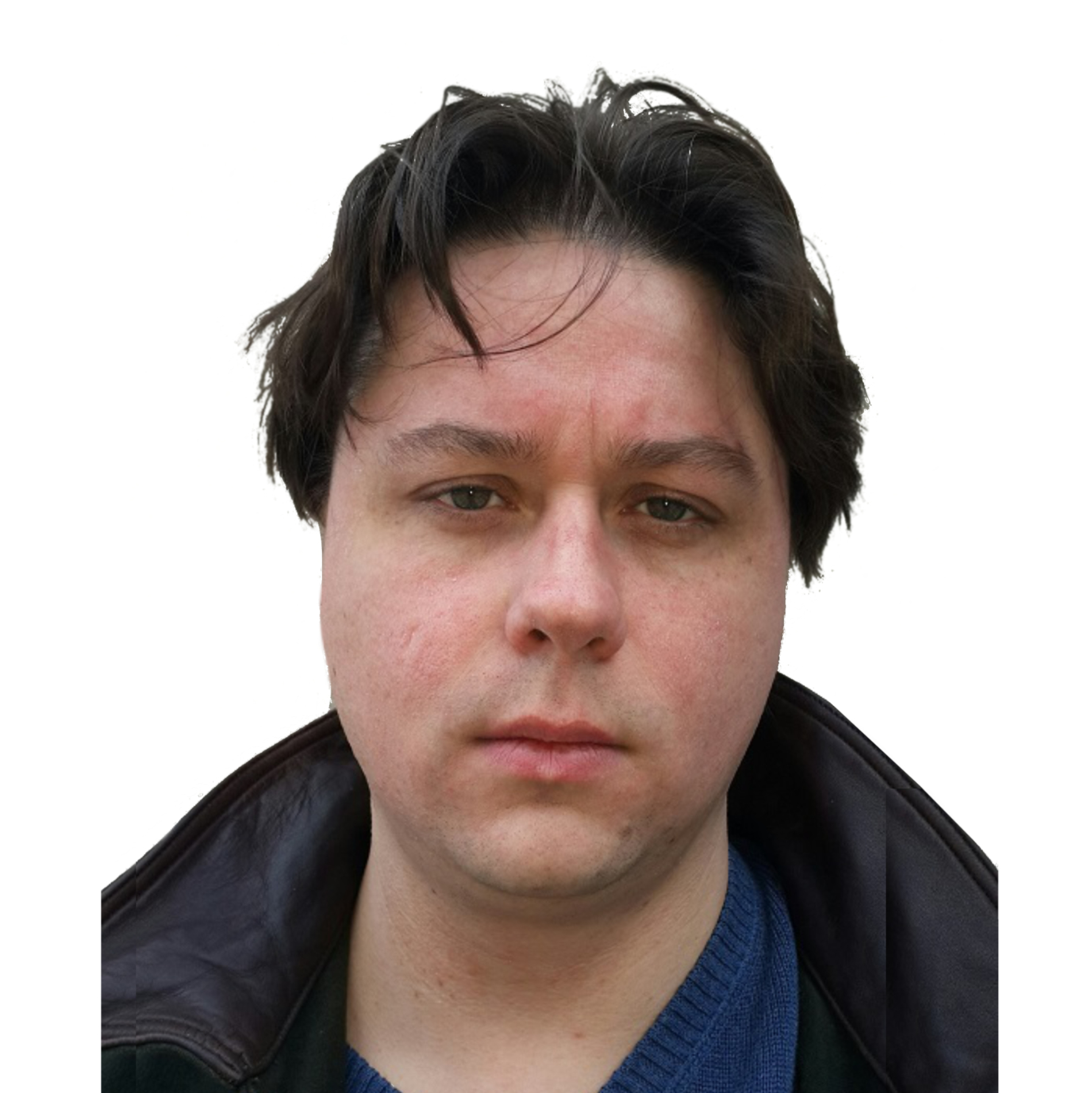}}]{Dmitri Moltchanov} received the MSc and Cand.Sc. degrees from the St. Petersburg State University of Telecommunications, Russia, in 2000 and 2003, and the Ph.D. degree from Tampere University of Technology, Finland, in 2006. Currently, he is an University Lecturer with the Laboratory of Electronics and Communications Engineering, Tampere University of Technology. He has (co)authored more than 150 publications on wireless communications, networking, the IoT applications, and applied queuing theory. He has taught more than 80 full courses on wireless and wired networking technologies, IoT systems, network modeling, and queuing theory. His research interests include 5G/6G systems, terahertz communications, ultra-reliable low-latency service, and industrial IoT applications.
\end{IEEEbiography}

\biographyspace

\begin{IEEEbiography}[{\includegraphics[width=1in,height=1.25in,clip,keepaspectratio]{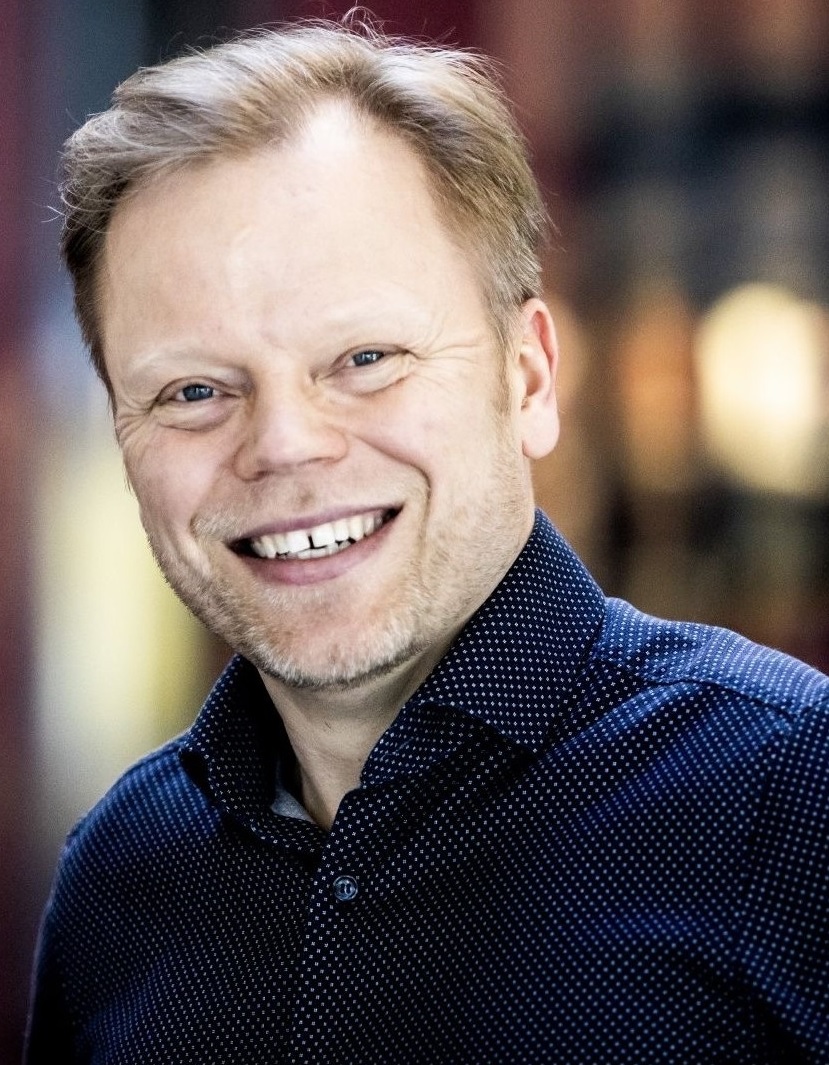}}]{Juho Pirskanen} is Head of Standardization and Research at Wirepas, Tampere, Finland, with over 25 years of experience in wireless radio technologies spanning 3G, LTE, WLAN, and 5G. He has been active in 3GPP and IEEE 802.11 standardization, and since 2018 has led contributions to ETSI DECT-2020 NR. He has (co-)authored numerous IEEE journal and conference papers on radio interface design, waveform processing, and positioning, as well as standards contributions. He holds an M.Sc. in Engineering from Tampere University.
\end{IEEEbiography}

\biographyspace

\begin{IEEEbiography}[{\includegraphics[width=1in,height=1.25in,clip,keepaspectratio]{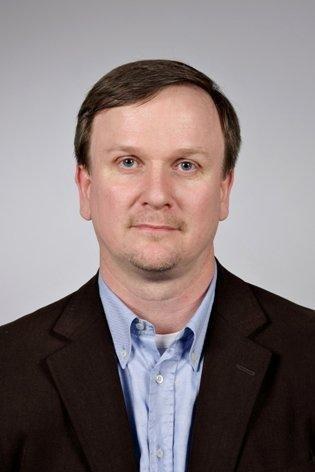}}]{Jussi Numminen} has worked with wireless standards for more than 25 years. He been working with standards in Nokia standardizing 4G and 3G technologies in 3GPP, and later cellular connectivity for ST Microelectronics, ST Ericsson and Ericsson. His current position is Vice president of Radio Strategy and IPR in Wirepas. Jussi is the Chair of the ETSI Technical Committee DECT, who is responsible for DECT-2020 NR standard development. This new standard fills a gap currently in the IoT connectivity environment; it addresses industrial applications requiring either or both massive scale (mMTC) and/or ultra-reliable low latency (URLLC), complementing today’s 5G technology in local networks.
\end{IEEEbiography}

\biographyspace

\begin{IEEEbiography}[{\includegraphics[width=1in,height=1.25in,clip,keepaspectratio]{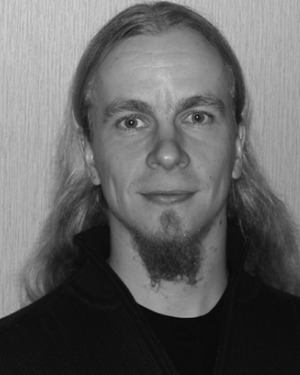}}]{Mikko Valkama} (Fellow, IEEE) received the M.Sc. (Tech.) and D.Sc. (Tech.) degrees (Hons.) in electrical engineering from the Tampere University of Technology, Tampere, Finland, in 2000 and 2001, respectively, where he is currently a Full Professor and the Unit Head of electrical engineering. His Ph.D. Dissertation was focused on advanced I/Q signal processing for wideband receivers: models and algorithms. In 2003, he was a Visiting Postdoctoral Research Fellow with the Communications Systems and Signal Processing Institute, San Diego State University, San Diego, CA, USA. His current research interests include radio communications, radio localization, and radio-based sensing, with particular emphasis on 5G and 6G mobile radio networks. He was a recipient of the Best Ph.D. Thesis Award of the Finnish Academy of Science and Letters for his Ph.D. Dissertation.
\end{IEEEbiography}

\end{document}